\documentclass[prd,aps,a4paper,floatfix,amsmath,amssymb,twocolumn]{revtex4-1}
\usepackage{graphicx,color,array,float,fullpage}

\newcommand{\br}{\nonumber \\}
\renewcommand{\S}{{\cal S}}
\newcommand{\C}{{\cal C}}
\newcommand{\T}{{\cal T}}
\newcommand{\rb}{{\bar r}}
\newcommand{\tb}{{\bar t}}
\newcommand{\rt}{{\tilde r}}

\newcounter{remembercounter}

\begin{document}


\title{Critical collapse of a rotating scalar field in $2+1$ dimensions}

\author{Joanna Ja\l mu\.zna}
\affiliation{ITFA and Delta ITP, Universiteit van Amsterdam, Science Park 904, 1098 XH Amsterdam, the Netherlands}
\affiliation{M. Smoluchowski Institute of Physics, Jagiellonian University, 30-348 Krak\'ow, Poland}
\author{Carsten Gundlach}
\affiliation{Mathematical Sciences, University of Southampton,
  Southampton SO17 1BJ, United Kingdom}
\date{15 February 2017, revised 16 March 2017}

\begin{abstract}
We carry out numerical simulations of the collapse of a complex rotating scalar field of the form $\Psi(t,r,\theta)=e^{im\theta}\Phi(t,r)$, giving rise to an axisymmetric metric, in 2+1 spacetime dimensions with cosmological constant $\Lambda<0$, for $m=0,1,2$, for four 1-parameter families of initial data. We look for the familiar scaling of black hole mass and maximal Ricci curvature as a power of $|p-p_*|$, where $p$ is the amplitude of our initial data and $p_*$ some threshold. We find evidence of Ricci scaling for all families, and tentative evidence of mass scaling for most families, but the case $m>0$ is very different from the case $m=0$ we have considered before: the thresholds for mass scaling and Ricci scaling are significantly different (for the same family), scaling stops well above the scale set by $\Lambda$, and the exponents depend strongly on the family. Hence, in contrast to the $m=0$ case, and to many other self-gravitating systems, there is only weak evidence for the collapse threshold being controlled by a self-similar critical solution and no evidence for it being universal.
\end{abstract}

\maketitle

\tableofcontents

\section{Introduction}

In the numerical and mathematical study of gravitational collapse,
massless scalar fields have often been used as a matter field. They
are simple, travel at the speed of light like gravitational waves, and
may also be of interest as fundamental fields. Similarly, the simplest
models of gravitational collapse are spherically symmetric, going back
to the key paper of Oppenheimer and Snyder \cite{OppenheimerSnyder} on
spherically symmetric collapse of dust to a Schwarzschild black hole.

Choptuik \cite{Choptuik1993} used the combination of massless scalar
field matter with spherical symmetry to spectactular effect,
initiating the study of critical phenomena in gravitational collapse:
the generic presence of universality, scaling and self-similarity in
the time evolution of initial data that are close to the threshold of
collapse. (Here we use the term collapse synonymously with black hole
formation from regular initial data).

The triad of universality, scaling and self-similarity was previously
familiar from critical phenomena at second-order phase transitions in
thermodynamics, understood in terms of renormalisation group
theory. Similarly, critical phenomena can be understood in terms of a
renormalisation group flow on the space of classical initial data in
general relativity that is at the same time a physical time evolution,
for suitable choices of the lapse and shift
\cite{GundlachLRR,symmetryseeking}. A novel feature in general
relativity is the appearance of discrete self-similarity (DSS), rather
than the continuous self-similarity (CSS) familiar elsewhere in
physics (such as fluid dynamics).

One obvious direction to go in from the work of Choptuik was to
generalise to axisymmetry. Abrahams and Evans \cite{AbrahamsEvans}
found scaling and DSS in the collapse of polarised axisymmetric vacuum
gravitational waves. These numerical results are widely believed to be
correct but have still not been verified independently.

With matter, axisymmetry is also the maximal symmetry in which rotating collapse
can be studied in 3+1 dimensions, leading to a Kerr black
hole. Moreover, in axisymmetry, angular momentum forms a
conserved current generated by the Killing vector field $K$,
\begin{equation}
j^a:={T^a}_bK^b \quad \Rightarrow \quad \nabla_aj^a=0.
\end{equation}
However, in 3+1 spacetime dimensions with axisymmetry, neither vacuum
gravitational waves or an axisymmetric massless scalar field
$\Phi$ can carry angular momentum. A simple way of seeing this for the
axisymmetric scalar field (for simplicity assumed to be real) is to note that
\begin{equation}
j^a=\nabla^a\Phi K^b\nabla_b\Phi-{1\over
  2}K^a(\nabla^b\Phi\nabla_b\Phi).
\end{equation}
The first term vanishes if the scalar field is itself axisymmetric,
and the second term is by definition tangent to any axisymmetric
slice, and so does not contribute to the Noether charge $J:=\int
j^a\,dS_a$.

A way of getting round that is to use a complex scalar field
  $\Psi=e^{im\varphi}\Phi$, where $\Phi$ is axisymmetric but now complex, and $m$ is
an integer. This results in an axisymmetric stress-energy tensor and
spacetime. Although using a complex scalar field appears to be a
natural choice, this particular ansatz introduces a pseudo-centrifugal
potential $2m/r$ (where $r$ is the cylindrical radius) into the wave equation even when the
angular momentum current vanishes identically, so the centrifugal
repulsion appears to be unrelated to angular momentum in a way that
appears to be atypical of intuitive ideas of the effect of angular
momentum in collapse. 

Choptuik, Hirschmann, Liebling and Pretorius
\cite{ChoptuikHirschmannLieblingPretorius} examined the case $m=1$ in
3+1 and found universality and scaling. The DSS critical solution is
distinct from the well-known one for $m=0$
\cite{Choptuik1993,critcont}. The $m=1$ critical solution is real (up
to a constant overall phase) and nonrotating and an attractor even for
rotating initial data, so that $J/M^2\to 0$ at the black-hole
threshold.

Critical collapse of a real scalar field in spherical symmetry in 2+1 was
investigated by Pretorius and Choptuik \cite{PretoriusChoptuik}, and
in more detail by us \cite{adscollapsepaper}. There are a number of
essential differences between 2+1 and all higher
dimensions. First, a negative cosmological constant is required to
form a black hole from regular data. This brings with it the existence
of a reflecting timelike outer boundary (at infinity). It also means that exactly
self-similar solutions cannot exist. Finally, because the mass in 2+1
dimensions is dimensionless, the black hole mass scaling cannot be
derived using a pure renormalisation group argument.

In \cite{adscollapsepaper} we investigated these issues in the
nonrotating ($m=0$) case with $\Phi$ real. We initially adopted as the definition of
supercritical initial data (in any given 1-parameter family of data)
that the Ricci scalar at the centre blows up without any preceding
minima and maxima. Fine-tuning to the critical parameter thus
identified, we found a universal critical solution that is
approximately (asymptotically on small spacetime scales) CSS inside
the lightcone of its (naked) singularity, but has a different symmetry
outside the lightcone. (The asymptotic form inside the lightcone had
previously been derived in closed form by Garfinkle
\cite{Garfinkle}). We also found the familiar and expected scaling of
the maximum of the Ricci scalar, 
\begin{equation}
\label{gammaexponentdef}
R_{\rm max}\sim (p_*-p)^{-2\gamma},
\end{equation}
for subcritical data \cite{GarfinkleDuncan}. In contrast to the scalar
field in higher dimensions, we actually saw scaling of the
values and locations of several maxima and minima of the Ricci
scalar. These must be features of the universal post-CSS subcritical
evolution, rather than the CSS critical solution itself.

We also found power-law scaling of the mass of the
earliest marginally outer-trapped surface (EMOTS), 
\begin{equation}
\label{deltadef}
M_{\rm EMOTS}\sim (p-p_*)^\delta. 
\end{equation}
We derived $\delta$ based on the interaction
of the critical solution outside its lightcone, the cosmological
constant and a single growing mode. Our argument relies on a
  technical conjecture, but is supported by the numerical observation that the
threshold value $p_*$ is the same for subcritical Ricci scaling and
supercritical mass scaling.

In the present paper, we investigate critical collapse for the
rotating axisymmetric complex scalar field ($m>0$ and/or $\Phi$ complex) in 2+1 dimensions.
This fills the gap in the $2\times 2$ table of models studied so far. As an additional motivation, axisymmetry in
2+1 dimensions reduces the field equations to partial differential
equations (PDEs) in only two coordinates $(t,r)$, even in the presence of
angular momentum, so that there is no extra computational cost
compared to spherical symmetry. Looking ahead to future work on
rotating fluid collapse in 2+1 dimensions, we have organised the
material so that Section~\ref{section:axisymmetry} and
Appendixes~\ref{appendix:gauge}-\ref{appendix:KerradS} hold for any
matter, while the rest of the paper is specific to rotating scalar
field matter.

\section{Axisymmetry with rotation}
\label{section:axisymmetry}

\subsection{Metric}

We consider 2+1-dimensional asymptotically anti-de Sitter (adS)
spacetimes with a rotational Killing vector $K=\partial_\theta$. For
clarity, we will refer to this symmetry as axisymmetry in general, but
as spherical symmetry in the absence of rotation, when there is an
additional reflection symmetry $\theta\to-\theta$.

In axisymmetry in 2+1 dimensions we make the metric ansatz
\begin{equation}
\label{trmetric}
ds^2=f(-dt^2+dr^2)+\rb^2(d\theta+\beta\,dt)^2,
\end{equation}
where
$f$, $\rb$ and $\beta$ are functions of $(t,r)$ only. To consider
asymptotically adS spacetimes, we rewrite this as
\begin{eqnarray}
\label{fdef}
f&:=&\frac{e^{2A}}{\C^2}, \\ 
\label{Rdef}
\rb&:=& e^B\ell\T, 
\end{eqnarray}
where $\ell$ is the length scale set by the cosmological constant
$\Lambda<0$ as
$\Lambda=:-1/\ell^2$, and where we have defined the shorthands 
\begin{equation}
\S:=\sin\left(\frac{r}{\ell}\right),
\C:=\cos\left(\frac{r}{\ell}\right),
\T:=\tan\left(\frac{r}{\ell}\right).
\end{equation}
The coordinate ranges are $-\infty<t<\infty$, $0\le r<\ell \pi/2$ and
$0\le \theta<2\pi$. It is helpful to keep in mind that in our
convention $\S=0$ at the centre $\rb=0$ of axisymmetry and $\C=0$
at the adS outer boundary $r=\ell \pi/2$. In this ansatz, $A=B=\beta=0$
represents the global adS spacetime.

These coordinates are a generalisation of those used in
\cite{PretoriusChoptuik,adscollapsepaper}. We show in
Appendix~\ref{appendix:KerradS} that the Kerr-adS solution can also be
expressed in these coordinates, so that these are
good coordinates for simulating rotating collapse. We discuss the
  remaining gauge freedom in Appendix~\ref{appendix:gauge}. The upshot
is that to fix the gauge completely we will impose $\beta=0$ at the outer boundary.

\subsection{Einstein equations}

The Einstein equations with a cosmological constant $\Lambda$ are
\begin{equation}
\label{Gab}
G_{ab}+\Lambda g_{ab}=4\pi T_{ab}
\end{equation}
in the units of \cite{PretoriusChoptuik,adscollapsepaper}, where
$G=1/2$ and $c=1$.  
In axisymmetry (with rotation), there are six independent components of
the Einstein equations. Two can be written as
wave equations for $A$ and $B$, namely
\begin{eqnarray}
-A_{,tt}+A_{,rr} \br
+C_3 -\frac{3}{4} C_4\gamma^2 +4\pi S_A&=&0,
\label{Awave} \\
- B_{,tt}+B_{,rr}+ \frac{2}{r}B_{,r}\br
+B_{,r}^2+B_{,r}\left(\frac{2}{\ell\S\C}
-\frac{2}{r}\right) - B_{,t}^2 \br 
+2C_3 +\frac{1}{2} C_4 \gamma^2 +4\pi S_B&=&0,
\label{Bwave}
\end{eqnarray}
where we have defined the shorthands 
\begin{equation}
C_3:=\frac{(1-e^{2A})}{\ell^2\C^2}, \quad
C_4:=\ell^2\S^2e^{2B-2A}. 
\end{equation}
We also have two constraint equations for $A$ and $B$, namely
\begin{eqnarray}
B_{,rr} +B_{,r}\left(B_{,r}-A_{,r}+\frac{1+\C^2} {\ell\S\C}\right)  
-\frac{A_{,r}}{\ell\S\C} \br
-A_{,t} B_{,t} +C_3  +\frac{1}{4}C_4\gamma^2 +4\pi S_{B'} =0, 
\label{Brconstraint}
\\
B_{,tr} +B_{,t}  \left( B_{,r}-A_{,r} +
\frac{\C}{\ell\S}\right) \br
-A_{,t} \left(B_{,r} +\frac{1}{\ell\S\C}\right)+4\pi S_{\dot B}=0.
\label{Btconstraint}
\end{eqnarray}
The last two Einstein equations (which become become trivial in spherical
symmetry) can be written as one evolution equation and one
constraint for
\begin{equation}
\label{gammadef}
\gamma:=\beta_{,r},
\end{equation}
namely
\begin{eqnarray}
\label{gammat}
J_{,t}+ 8\pi\rb S_{\dot\gamma} &=& 0, \\
\label{gammar}
J_{,r}+ 8\pi\rb S_{\gamma'} &=& 0, 
\end{eqnarray}
where we have defined the shorthand
\begin{equation}
\label{Jdef}
J:={\rb^3 \gamma\over f}.
\end{equation}
We show in Appendix~\ref{appendix:KerradS} that in vacuum  
 $J$ is the angular momentum parameter of the BTZ metric.
We have also introduced the following shorthands for the source terms
of the six Einstein equations:
\begin{eqnarray}
S_A&:=&-{e^{2A-2B}\over\ell^2\S^2} T_{\theta\theta} \\
S_B&:=&T_{tt}-T_{rr}-2\beta T_{t\theta}+4\beta^2T_{\theta\theta} \\
S_{\dot B}&:=& T_{tr}-\beta T_{r\theta} \\
S_{B'}&:=&T_{tt}-2\beta T_{t\theta}+4\beta^2T_{\theta\theta} \\
S_{\dot\gamma}&:=& T_{r\theta}, \\
S_{\gamma'}&:=& T_{t\theta}-\beta T_{\theta\theta}.
\end{eqnarray}

\subsection{Apparent horizon}

A marginally outer-trapped surface (MOTS) in axisymmetry is
given by 
\begin{equation}
\label{MOTScrit}
g_+:=\left({\partial_t+\partial_r}\right)\ln\rb=B_{,t}+B_{,r}+{1\over\ell\S\C}=0.
\end{equation}
The curve $g_+=0$ in the $tr$-plane defines the apparent horizon
(AH). An isolated horizon (IH) is a piece of the AH that is null.
The apparent horizon is spacelike, timelike or null if the product
$g_{+,v}g_{+,u}$ is positive, negative or zero. 

\subsection{Quasilocal angular momentum}
\label{section:quasilocangmom}

Consistently with (\ref{Jdef},\ref{gammar}), we define the quasilocal angular momentum
\begin{equation}
\label{Jintegral}
J(t,r)=8\pi \int_0^r \omega \sqrt{\gamma}\,dr',
\end{equation}
where 
\begin{equation}
\label{omegadef}
\omega:=-j^an_a=-T^{ab}n_aK_b ={T^t}_\theta\sqrt{f}=-{S_{\gamma'}\over \sqrt{f}}
\end{equation}
is the angular momentum density per unit volume of space, and $n^a$
and $\gamma$ are the future-pointing unit normal and volume element on
slices of constant $t$.  $J$ is the conserved quantity related to
$j^a$, and is therefore gauge-invariant and independent of the time
slice on which we have integrated from the centre out to the point
$(t,r)$.

Because mass in 2+1 spacetime dimensions is dimensionless,
$\omega$ has units of $1/{\rm length}$, and $J$ has units of length.
The factor of $8\pi$ has been inserted so that $J$ coincides with the
expression for $J$ in Kerr-adS spacetime. It is therefore constant in
vacuum and reduces to the BTZ angular momentum.

\subsection{Quasilocal mass candidates}

A possible quasilocal mass expression is the local BTZ mass parameter
\begin{equation}
M_{\rm BTZloc}(t,r):={\rb^2\over\ell^2}+{J^2\over 4\rb^2}-(\nabla \rb)^2.
\end{equation}
This is a scalar, and reduces to the constant BTZ mass in
vacuum. However, we will see that, at least for the complex scalar
field considered here, its mass aspect $M_{{\rm BTZloc},r}$ may become
negative.

Alternatively, we could extend the 2+1 dimensional Hawking mass from
spherical symmetry \cite{PretoriusChoptuik}
\begin{equation}
M_{\rm H}(t,r):={\rb^2\over\ell^2}-(\nabla \rb)^2
\end{equation}
to axisymmetry. The mass aspect $H_{{\rm H},r}$ is non-negative for
rotating scalar field matter. However, $M_{\rm H}$ is not constant in
the Kerr-adS solution. 

On the horizon of a stationary black hole, or more generally on any
isolated horizon (IH) characterised by $|\nabla\bar r|^2=0$, $M_{\rm
  BTZloc}$ reduces to the BTZ mass $M$, while the generalised Hawking
mass reduces to the irreducible mass:
\begin{equation}
\left.M_{\rm H}\right|_{\rm IH}={\rb^2\over\ell^2}
={1\over2}\left(M+\sqrt{M^2-{J^2\over\ell^2}}\right)=M_{\rm irr}.
\end{equation}
(In any dimension, the irreducible mass is uniquely defined by the
requirements that $dM-\Omega\,dJ>0$ if and only if $dM_{\rm irr}>0$
and $M_{\rm irr}=M$ for non-rotating black holes \cite{Wald}). 

In the following, we exclusively use $M:=M_H$ as our quasilocal mass.

\section{Rotating scalar field matter}

\subsection{Field equations}

The stress-energy tensor for a minimally coupled
massless complex scalar field $\Psi$ is
\begin{equation}
\label{Tab}
T_{ab}={1\over 2}\left(\Psi_{,a} \Psi^*_{,b} +\Psi^*_{,a}
\Psi_{,b} - g_{ab}g^{cd}\Psi_{,c} \Psi^*_{,d}\right).
\end{equation}
This is conserved, $\nabla_aT^{ab}=0$, if and only
if $\Psi$ obeys the wave equation $\nabla_a\nabla^a\Psi=0$. 

We make the axisymmetric rotating complex scalar field ansatz
\begin{equation}
\label{Psidef}
\Psi= e^{im\theta}\Phi, \quad \Phi=:\S^m (\phi+i\psi),
\end{equation}
where $\phi$ and $\psi$ are real and independent of $\theta$, and $m$
is an integer. Without loss of generality we set $m\ge 0$ from now
on. In this ansatz, regularity of $\Psi$ requires $\phi$ and $\psi$ to
be even and regular [and hence generically $O(1)$] in $r$ at the
origin $r=0$. We have chosen the regularisation factor $\S^m$ in (\ref{Psidef})
rather than $r^m$ or $\rb^m$ because this gives rise to the simplest
form of the field equations.

With the first-order variables
\begin{eqnarray}
V&:=&\phi_{,t}+m\beta\psi, \quad X:=\phi_{,r},  \\
W&:=&\psi_{,t}-m\beta\phi, \quad Y:=\psi_{,r},
\end{eqnarray}
the coupled wave equations for $\phi$ and $\psi$ are are
\begin{eqnarray}
\label{phiwave}
-V_{,t}+X_{,r}+\frac{2m+1}{r}X && \br
+C_1X-m\beta W -B_{,t}V +C_2\phi &=&0, \\
-X_{,t}+V_{,r}-m(\beta Y+\gamma\psi)&=&0, 
\end{eqnarray}
and 
\begin{eqnarray}
\label{psiwave}
-W_{,t}+Y_{,r}+\frac{2m+1}{r}Y && \br
+C_1Y+m\beta V -B_{,t}W +C_2\psi &=&0, \\
-Y_{,t}+W_{,r}+m(\beta X+\gamma\phi)&=&0,
\end{eqnarray}
where we have introduced the shorthands
\begin{eqnarray}
C_1&:=&\left(\frac{2m}{\ell\T}+\frac{1}{l\S\C}-\frac{2m+1}{r}\right) +B_{,r}, \\
C_2&:=&m\frac{B_{,r}}{\ell\T}+m^2\frac{\C^2-e^{2A-2B}}{\ell^2\S^2}
\end{eqnarray}

Looking at the ensemble of all field equations, the first lines of (\ref{phiwave},\ref{psiwave}),
(\ref{Awave},\ref{Bwave}) and (\ref{Brconstraint},\ref{Btconstraint})
represent their principal parts, where we must consider terms of the
type $B_{,r}/r$ and $X/r$ as principal in analysing well-posedness and
numerical stability. We have already eliminated all terms of the type
$\phi/r^2$, which otherwise we would also consider principal, by
introducing the factor $\S^m\simeq r^m$ in (\ref{Psidef}).

The source terms for the Einstein equations with scalar field matter are
\begin{eqnarray}
\left.\begin{array}{c}S_A\\S_{B'}\end{array}\right\}
&=&{1\over2} \S^{2m}[(X^2+Y^2)\mp(V^2+W^2)] \br 
&& +m^2{\S^{2m-2}\over2\ell^2}\left(\C^2\mp
e^{2A-2B}\right)(\phi^2+\psi^2)\br 
&& + m\frac{\C\S^{2m-1}}{\ell}(X\phi+Y\psi), \\
S_B&=&m^2{\S^{2m-2}\over\ell^2}e^{2A-2B}(\phi^2+\psi^2),\\
S_{\dot B}&=& \S^{2m}(VX+WY)  \br 
&& +m {\C\S^{2m-1}\over\ell}(V\phi+W\psi), \\
S_{\dot\gamma}&=&m\S^{2m}(Y\phi-X\psi), \\
S_{\gamma'}&=&m\S^{2m}(W\phi-V\psi).
\end{eqnarray}

For $m=0$ and $\beta=0$, the Einstein equations reduce to Eqs.~(6-9)
of \cite{PretoriusChoptuik}, but with two copies of the scalar field.
For $m>0$, we can consistently restrict solutions to the class of
real, non-rotating solutions where $\psi$ and $\beta$, and hence $W$,
$Y$, $\gamma$ all vanish.

As a curvature diagnostic we use the Ricci scalar
\begin{eqnarray}
R&=&-{6\over\ell^2}+8\pi\C^2\S^{2m-2}e^{-2A}\Bigl[ \br
&&  \S^2(X^2+Y^2-V^2-W^2)+{2m\S\C\over\ell}(X\phi+Y\psi) \br
&&+{m^2\over\ell^2}(e^{2A-2B}+\C^2)(\phi^2+\psi^2)\Bigr].
\end{eqnarray}
Note that this expression vanishes at $r=0$ except for $m=0$ (with $V^2+W^2$
contributing) and $m=1$ (with $\phi^2+\psi^2$ contributing). 

For the rotating scalar field, we use the diagnostic $\omega$ or
\begin{equation}
\bar\omega:={\omega\over m}=e^{-A}\C\S^{2m}(W\phi-V\psi),
\end{equation}
where $\bar\omega$ is defined also for $m=0$. If the complex scalar field was
coupled to an electromagnetic field, $\bar\omega$ would be the electric
charge density of the scalar field. It is an artifact of our
ansatz for $\Psi$ that its angular momentum density $\omega$ is simply
equal to its ``charge density'' $\bar\omega$, times the integer $m$.

\subsection{Symmetries and boundary conditions}

At $r=0$, the boundary conditions follow from the fact that $A$, $B$,
$\beta$, $\phi$, $\psi$, $V$, $W$ are even in $r$ and generically
$O(1)$ (and so obey Neumann boundary conditions), and $\gamma$, $X$
and $Y$ are odd and generically $O(r)$ (and so obey Dirichlet boundary
conditions). There is one additional geometric regularity condition,
namely the absence of a conical singularity at $r=0$, or
\begin{equation}
A(0,t)-B(0,t)=0.
\end{equation}
Together, all these conditions are equivalent to the standard
requirement that the metric and scalar fields must be analytic functions at
$x=y=0$ when expressed in the Cartesian coordinates $x=r\cos\theta$
and $y=r\sin\theta$. They are of course compatible with the field
equations.

At the timelike adS infinity, regularity of (\ref{Awave},\ref{Bwave})
requires $A,\phi,\psi\sim z^2$ and $B\sim z$, where we have defined
the shorthand $z:=r-\ell\pi/2$,. Hence $A$, $\phi$ and $\psi$ obey
both Dirichlet and Neumann boundary conditions, and $B$ obeys
Dirichlet boundary conditions. The first-order auxiliary variables
$V,W,X,Y$ therefore all vanish at the adS boundary.  As already
discussed, we also impose the gauge boundary conditon (\ref{gaugeBC}).

It is compatible with the field equations to assume that $A$, $B$,
$\beta$, $\phi$, $\psi$ are even functions of $z$ (as well as even
functions of $r$). This is true because of the way $r$ appears in the
field equations only through $\S$ and $\C$. With this assumption, the
variables $A,B,\beta,\phi,\psi$ are even and $X,Y,\gamma$ are odd,
about both boundaries. In addition $A,\phi,\psi,V,W$ also vanish at
$z=0$. However, unlike $r=0$, $z=0$ is not an interior point of the
spacetime, and so this symmetry does not follow from regularity
alone. Rather, it can be imposed as a consistent restriction of the
solution space. In the following, we always make this assumption (as we already did in \cite{adscollapsepaper}).

\subsection{Apparent horizon}

Recall that the apparent horizon is spacelike, timelike or null if the quantity
$g_{+,v}g_{+,u}$ is positive, negative or zero. 
The two factors of this expression, using
the Einstein equations, are 
\begin{eqnarray}
g_{+,v}&=&-4\pi\S^{2m}\Biggl[\left(V+X+{m\phi\over\ell\T}\right)^2
\br && +\left(W+Y+{m\psi\over\ell\T}\right)^2\Biggr],
\\ g_{+,u}&=&-{2e^{2A-2B}\over\ell^2}
\Biggl[{e^{2B}\over\C^2}-{1\over 4}e^{4B-4A}\ell^2\S^2\gamma^2
\br && -2\pi m^2\S^{2m-2}(\phi^2+\psi^2)\Biggr].
\end{eqnarray}
In the $m=0$ case, $g_{+,u}<0$, and $g_{+,v}\le 0$ with equality only at
$r=0$ or where $\Psi_{,v}=0$. Hence we recover the result
\cite{adscollapsepaper} that the AH is null (becoming an IH) where
$\Psi_{,v}=0$ and spacelike elsewhere. For $m>0$, $g_{+,v}\le 0$ still
holds but $g_{+,u}$ can now become positive for $\phi^2+\psi^2$
sufficiently large, even in the absence of rotation. Hence the AH can
become timelike in the presence of matter. However,
  $\Psi_{,v}=0$ (no infalling matter) still implies that the AH is
  null (becomes an IH).

\subsection{Quasilocal mass}

The mass aspect $M_{H,r}$ for rotating scalar field matter is
\begin{eqnarray}
M_{H,r}&=&f_1+f_2, \\
f_1&:=&g_-P[(V+\hat X)^2+(W+\hat Y)^2] \br
&&+g_+P[(V-\hat X)^2+(W-\hat Y)^2] \br 
&&4\pi m^2g_r \S^{2m} (\phi^2+\psi^2), \\
f_2&:=&{1\over 2}e^{4B-4A}\ell^4\S^4\gamma^2,
\end{eqnarray}
where $g_r:=(\ln\rb)_{,r}$ and we have defined the positive definite factor
\begin{equation}
P:= 2\pi\ell^2e^{2B-2A}\S^{2m+2}
\end{equation}
and the shorthands
\begin{equation}
\hat X:=X+{m\phi\over\ell\T}, \quad \hat Y:=Y+{m\psi\over\ell\T}.
\end{equation}
Outside the AH, we have $g_+,g_-,g_r>0$, and hence $M_{H,r}$ is
positive outside the AH. In particular, we recover the corresponding
result for the case $m=0$ stated in \cite{adscollapsepaper}. However,
for $m>0$, $M_{H,r}$ vanishes in vacuum only if $\gamma=0$.

By contrast $M_{{\rm BTZloc},r}$ vanishes in vacuum, even with
rotation, but is not positive definite. It can be written as
\begin{eqnarray}
M_{{\rm BTZloc},r}&=&f_1-f_3, \\
f_3&:=&2mP\gamma(W\phi-V\psi).
\end{eqnarray}
Note that
\begin{equation}
\left({J^2\over 4\rb^2}\right)_{,r}=-f_2-f_3,
\end{equation}
where the indefinite term $f_3$ comes from $(J^2)_{,r}$.

\section{Numerical method}

\subsection{Einstein equations}

We use a numerical grid in $(t,r)$, with Courant factor $\Delta
r/\Delta t=1/4$, so that taking every fourth time slice we trivially
obtain a double null grid in $(u,v)$. (In
  \cite{adscollapsepaper}, we used $\Delta r/\Delta t=1/64$, but this
  is unnecessary for stability, and we have checked that there is no
  significant difference in results.) However, for all numerical
purposes, we are doing a Cauchy evolution. We set data on $t=0$, and
evolve forward in $t$, with regularity boundary conditions at $r=0$
and $z=0$. We use standard fourth-order central finite-differencing
in $r$ (except for the principal part of the wave equation, see below), and fourth-order Runge-Kutta in $t$.

The constraints (\ref{Btconstraint},\ref{Brconstraint}) can be solved
on a time slice either as coupled ordinary differential equations
in $r$ for $B$ and $B_{,t}$, or as coupled algebraic equations
for $A_{,t}$ and $A_{,r}$ (and then by integration in $r$ for
$A$). Generalising the approach of \cite{PretoriusChoptuik}, we
  make the gauge choice $B=B_{,t}=0$ at $t=0$, fix initial data
  $\phi$, $\psi$, $X=\phi_{,r}$, $Y=\psi_{,r}$, $V$ and $W$, and
  iteratively solve the constraints for $A$, $A_{,t}$ and $\gamma$. We
  then obtain $\beta$ from $\gamma$ by integration. During the
  evolution for $t>0$, we then solve the wave equations for $A$ and
  $B$, as well as the wave equations for $\phi$ and $\psi$.

At $t=0$, and then at each time step, we find $\gamma$ by integrating
(\ref{gammar}) outwards, and then $\beta$ by integrating
(\ref{gammadef}) inwards. We either do the same at each time step (and
at each Runge-Kutta substep), or we use (\ref{gammat}) to evolve
$\gamma$ by integration, and again obtain $\beta$ by integrating
(\ref{gammadef}) inwards.

To obtain the correct behaviour $\gamma\sim \S^{2m-1}$ as $r\to 0$, we
write (\ref{gammar}) as
\begin{equation}
\gamma(r)={8\pi m\over \ell^2 C_5}\int_0^r  \frac{e^B}{\C^2}
(W\phi-V\psi){d(\S^{2m+2})\over 2m+2},
\end{equation}
and apply Simpson's rule for unequally spaced points $\S^{2m+2}$ to the integral with cubic spline interpolation for the middle points.  This
explicit expression assumes that $B$ is given, so this integration has
to be carried out at each Runge-Kutta substep in time. At $t=0$, we
set $B=0$.

With $i=N$ the outer boundary grid point, so that $A_N=B_N=\beta_N=0$,
we update the point $i=N-1$ by using the four grid points $N-2$,
$N-3$, $N-4$ and $N-5$ to fit the polynomial
$A=A_2z^2+A_4z^4+A_6z^6+A_8z^8$ where $z:=r-\ell\pi/2$ and evaluate it
at gridpoint $N-1$, and similarly for $\phi,\psi$, which we also
assume to be even in $z$ and which also vanish. For $B$, which is even
but does not vanish at $z=0$, we fit $B=B_0+B_2z^2+B_4z^4+B_6z^6$
instead.

\subsection{Wave equation}

The principal part of our wave equation for $\phi$ (and similarly
for $\psi$), is
\begin{eqnarray}
\label{Xprincipal}
-V_{,t}+X_{,r}+\frac{p}{r}X &=&0, \\
\label{Vprincipal}
-X_{,t}+V_{,r}&=&0,
\end{eqnarray}
where $p:=2m+1$.  Even though X is $O(r)$ and even, so that $X/r$ is
regular, a naive finite differencing of this linear wave equation is
known to suffer from numerical instabilities that quickly become
unmanageable with increasing $p$ (in our case, increasing $m$). A
stable finite differencing for arbitrarily large $p$ based on
summation by parts (SBP) has been given in \cite{lwaveSBP}.

In Appendix~\ref{appendix:SBP} we give explicit formulas for the
method we use, the SBP42 method for a centred grid. In the continuum
limit in time, this finite differencing scheme in space can be proved
to be stable for all positive integers $p$ in a discrete energy norm
that mimics the continuum energy for this wave equation. One can
prove convergence to fourth order in the discrete energy norm before
the wave interacts with the outer boundary, and to second-order
afterwards. No numerical viscosity is required. Numerical experiments
described in \cite{lwaveSBP} in fact still show third-order
convergence after interaction with the boundary, and this is true
not only in the energy norm but pointwise.

\subsection{Apparent horizon and EMOTS}
On each time slice, we locate the apparent horizon by finding up to
four zeros $r_{A,B,C,D}$ of $g_+(r)$. When we plot $r_{A,B,C,D}$
against $t$ this gives us the shape of the AH curve $t_{AH}(r)$. We
allow for four points in case the AH curve is W-shaped -- this did
happen for $m=0$. Two points are sufficient if it is V-shaped -- we
have always found this for $m>0$. Let $t_1$ be the first time step for
which we find nontrivial values $r_{A,B}$. We then approximate $t_{\rm
  EMOTS}=t_1$, and $r_{\rm EMOTS}=(r_A+r_B)/2$.

\section{Numerical results}

\subsection{Initial data}

In order to separate the initial implosion of a wave packet from its
reflections at the adS boundary as clearly as possible we make the
initial data as ingoing as possible 
as is compatible with $X$ being odd and $\phi$ and $V$ being even in
$r$. Hence we set
\begin{eqnarray}
\phi(0,r)&=&f_\phi(r)+f_\phi(-r), \\
X(0,r)&=&f_\phi'(r)-f_\phi'(-r), \\
V(0,r)&=&f_\phi'(r)+f_\phi'(-r),
\end{eqnarray}
and similarly for $\psi$, $Y$ and $W$. 
We rely on the initial data being very small at the
outer boundary for them to trivially obey the boundary conditions
there. 

We have investigated four
families of initial data of this type.  In family A we take $f_\phi$
to be a Gaussian with centre $r_{0\phi}=0.2$, width $\sigma_r=0.05$
and amplitude $p$, and we set $f_\psi=0$, so that these data are real
and non-rotating. In family B we set $f_\phi$ as we do in family A,
with $r_{0\phi}=0.2$, and $f_\psi$ with the same amplitude and width
but centre at $r_{0\psi}=0.25$.  In family C, we set both $f_\phi$ and
$f_\psi$ to be Gaussians with the same parameters as in family A, but
multiply $f_\phi$ by $\cos\omega r$ and $f_\psi$ by $\sin\omega r$,
with $\omega=200$. In family D we set $f_\phi$ and $f_\psi$ as
  Gaussians with the same amplitude, and again width $\sigma_r=0.05$,
  but now with centres $r_{0\phi}=0.05$ and $r_{0\psi}=0.1$. To
fine-tune to the threshold, in each family of initial data we vary the amplitude $p$.

We evolve until $t=2$, that is two light-crossing times, or until an
EMOTS and then a singularity forms, on a grid with
$N=1000$ points, and $N=2000$ for family D. For nonrotating data, we
optionally excise the central singularity when it forms, using the
simple causal structure of our coordinates. (When $\beta\ne 0$ this is
not possible because we solve $\beta=\int \gamma\,dr$.)

The critical amplitudes, critical exponents, and ranges over which we
see scaling are summarised in Table~\ref{table:fits}. 

Throughout this
paper we use the shorthand terminology of \cite{adscollapsepaper},
where ``sub$n$'' denotes sub-critical data ($p<p_*$) with
$\ln(p_*-p)=n$, and ``super$n$'' denotes supercritical data ($p>p_*$)
with $\ln(p-p_*)=n$, so data with larger $n$ are closer to critical.
In contrast to the non-rotating case treated in
\cite{adscollapsepaper}, we will see that for $m>0$ there are distinct
critical values of $p$ for the scaling of the maximum of the Ricci
scalar $R$ and local angular momentum density $\omega$ on the one
hand, and for the scaling of the EMOTS mass $M$ on the other. We will
use $p_*$ and ``sub$n$'' for the Ricci and angular momentum
scaling, while for mass scaling we introduce 
$p_{*M}$ and ``superM$n$''. In particular, for $m>0$
(\ref{deltadef}) is replaced by
\begin{equation}
\label{deltadefbis}
M_{\rm EMOTS}\sim (p-p_{*M})^\delta. 
\end{equation}

\begin{table*}
\setlength{\tabcolsep}{6pt}
\begin{tabular}{ll|llllll}
family & $m$ & $p_*$  & $p_{*M}$ & $2\gamma$  & $\delta$
& sub$n$ & superM$n$ \\
\hline
A &0& 0.13305923 & 0.13305923 & 2.36 & 0.69 & [-25,-5] & [-25,-15] \\
B&0& 0.08462225 & 0.08462225 & 2.46 & 0.66 & [-18,-4] & [-16,-4]  \\
C &0& 0.01356158 & 0.01356158 & 2.27 & 0.67 & [-18,-6] & [-18,-5]  \\
D &0& 0.183241&0.183241& 2.53 & 0.54 & [-13,-4] & [-13,-2]\\
\hline
A &1& 0.576 & 0.4399 & 3.73 & 0.42 & [-6.5,-2] & [-5,-2.5] \\
B&1& --- & 0.257 & --- & 0.36  & --- & [-4.4,-2]\\
C &1& 0.066957 & --- & 9.54 & ---  & [-8,-3.5] & --- \\
D &1& 1.1269357 & 1.1 & 3.93 & 0.03 & [-13,-4] & [-7,-4]\\
\hline
A &2& 1.632 & 1.395 & 3.11 & 0.1 & [-6,-0.5] & [-4.5,-2] \\
B&2& 0.871 & 0.74 & 1.98 & 0.16 & [-7,-1] & [-3.5,-2] \\
C &2& 0.19103 & 0.132 & 5.36 &0.623 & [-9,-3] & [-5,-3.5] \\ 
D &2& 4.682 & 4.544 & 1.93 & 0.05 & [-7,-2] & [-6,-3] \\
\end{tabular}
\caption{Values of critical amplitudes, critical exponents, and
  approximate ranges of $\ln|p-p_*|$ for which we observe scaling, for
  $m=0,1,2$ and four different families of initial data. $p_*$ is the
  critical value of $p$ for Ricci and angular momentum density
  scaling, and $\gamma$ is the corresponding critical exponent, see
  (\ref{gammaexponentdef}) and (\ref{omegagamma}). $p_{*M}$ is the critical
  value for EMOTS mass scaling, and $\delta$ the corresponding
  critical exponent, see (\ref{deltadefbis}).}
\label{table:fits}
\end{table*}

\subsection{Evidence for self-similarity and subcritical scaling}
\label{section:self-similarityevidence}

We start by looking for direct evidence that, for sufficient
fine-tuning of the parameter $p$ to some threshold value $p_*$, the
time evolution goes through a universal (for given $m$) self-similar
phase. A continuously self-similar and axisymmetric metric can be
written in coordinates $(T,x,\theta)$ adapted to both symmetries as
\begin{equation}
\label{Txdef}
g_{\mu\nu}(T,x,\theta)=e^{-2T}\bar g_{\mu\nu}(x).  
\end{equation}
It follows
that during this hypothetical self-similar phase
\begin{equation}
\label{RCSS}
R(t,0)\simeq a(t_{0*}-t_0)^{-2},
\end{equation}
where $R$ is the Ricci scalar, $t_0=t_0(t)$ the proper time at the
origin, $t_{0*}$ is a family-dependent accumulation time (obtained by
fitting) and $a$ a universal dimensionless constant. (See
\cite{GundlachLRR} for a general argument and
\cite{adscollapsepaper} for a detailed discussion of the case of
spherical symmetry in 2+1.) To look for this behaviour, we plot $\ln
|R(t_0,0)|$ against $\ln |t_{*0}-t_0|$ and adjust the parameter
$t_{*0}$ to optimise the linear fit.

This self-similar phase ends when the/a growing mode of the critical
solution has reached some nonlinearity threshold, and this must happen
at
\begin{equation}
\label{t0nonlin}
A|p-p_*|^{-\gamma}(t_{0*}-t_0)\simeq 1, 
\end{equation}
where $\gamma$ is the critical exponent, and $A$ is a family-dependent
constant. To the extent that we can neglect the infall of further
matter (which for $\Lambda<0$ is often not true), the subsequent
evolution is no longer self-similar but is still universal up to an overall
spacetime scale, so that in partciular
\begin{equation}
\label{RpostCSS}
|R(t,0)|\simeq
A^2|p-p_*|^{-2\gamma}f_\pm\left[A|p-p_*|^{-\gamma}(t_0-t_{0*})\right],
\end{equation}
where $A$ is the same as in (\ref{t0nonlin}), and so depends on
the family of initial data, but the two dimensionless functions
$f_\pm$ (one for $p>p*$ and one for $p<p_*$) are universal. $f_+$
obviously blows up.

\begin{figure}[!htb]
\centering
\includegraphics[scale=0.3, angle=270]{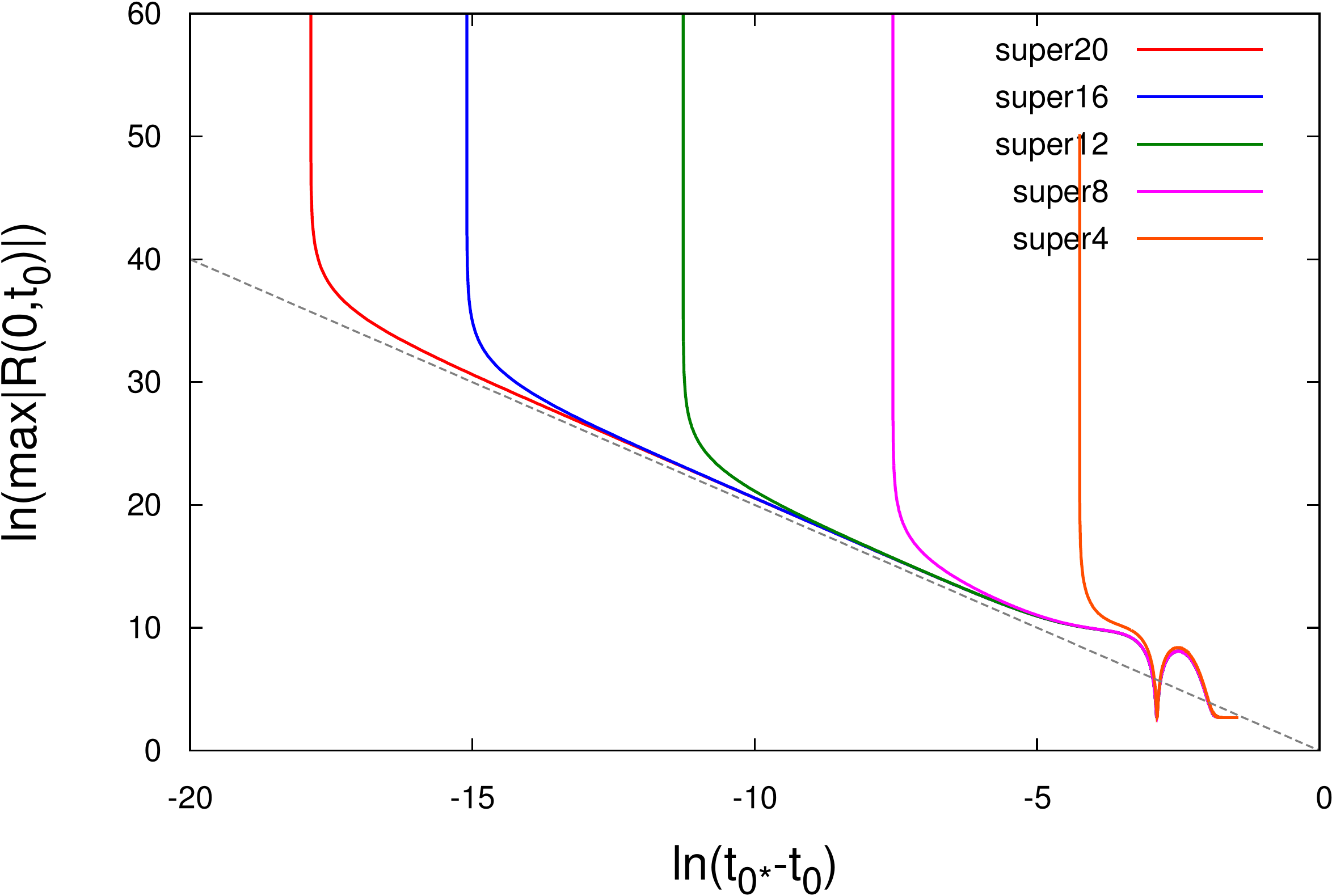}
\includegraphics[scale=0.3, angle=270]{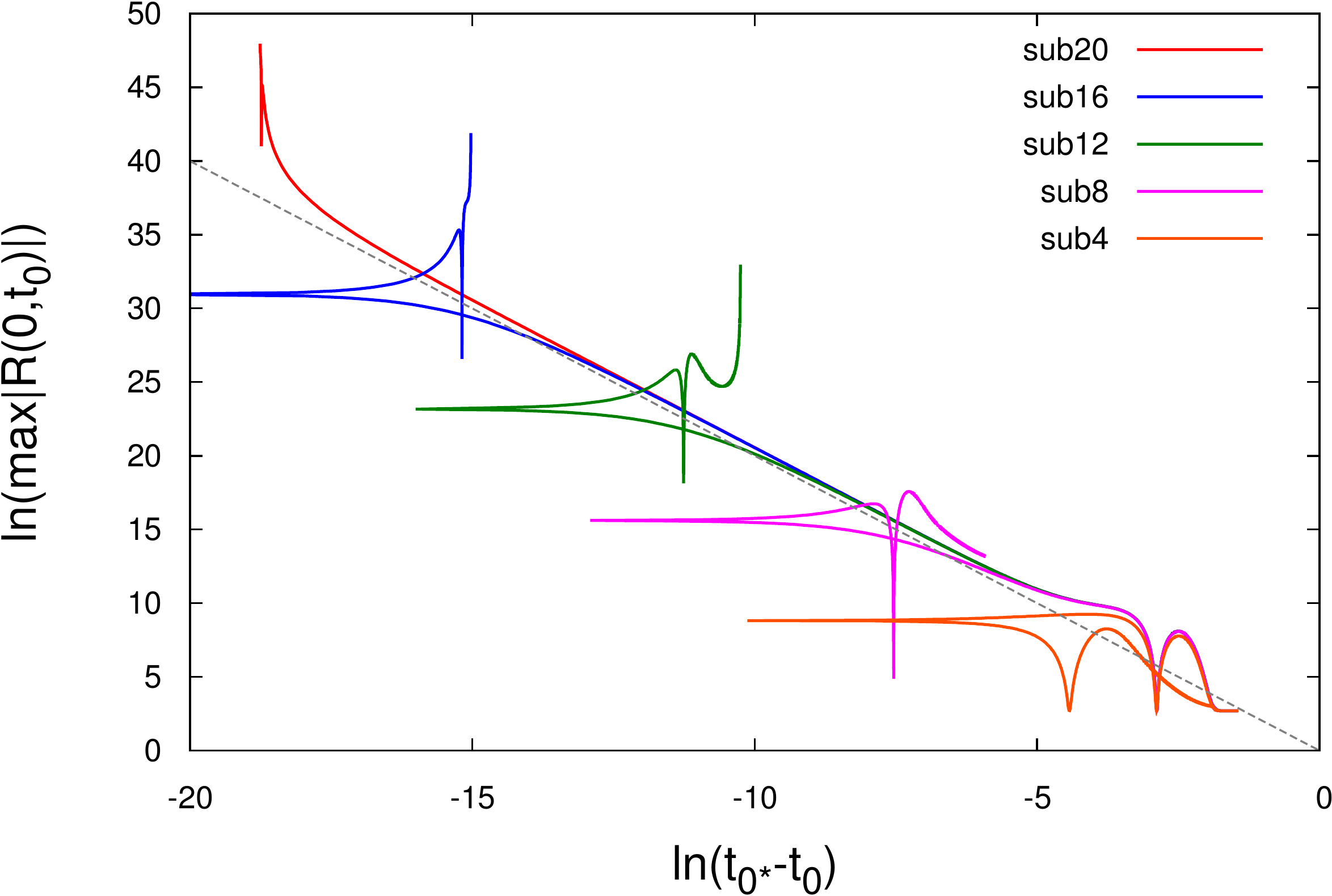}
\caption{$m=0$ A data: $\ln|R(t,0)|$ against $\ln |t_{*0}-t_0|$ for
  a few supercritical (top) and subcritical (bottom)
  evolutions. These correspond to data investigated in
  \cite{adscollapsepaper} and \cite{PretoriusChoptuik}. The
    grey line represents
    $R(t,0)=|t_{*0}-t_0|^{-2}$, indicating that indeed the spacetime
    is CSS near the centre. In the labels, ``sub$n$'' means that
    $-\ln(p_*-p)=n$, and ``super$n$'' means $-\ln(p-p_*)=n$.}
\label{figure:m0_Ar_ricci_t0}
\end{figure}

\begin{figure}[!htb]
\centering
\includegraphics[scale=0.3, angle=270]{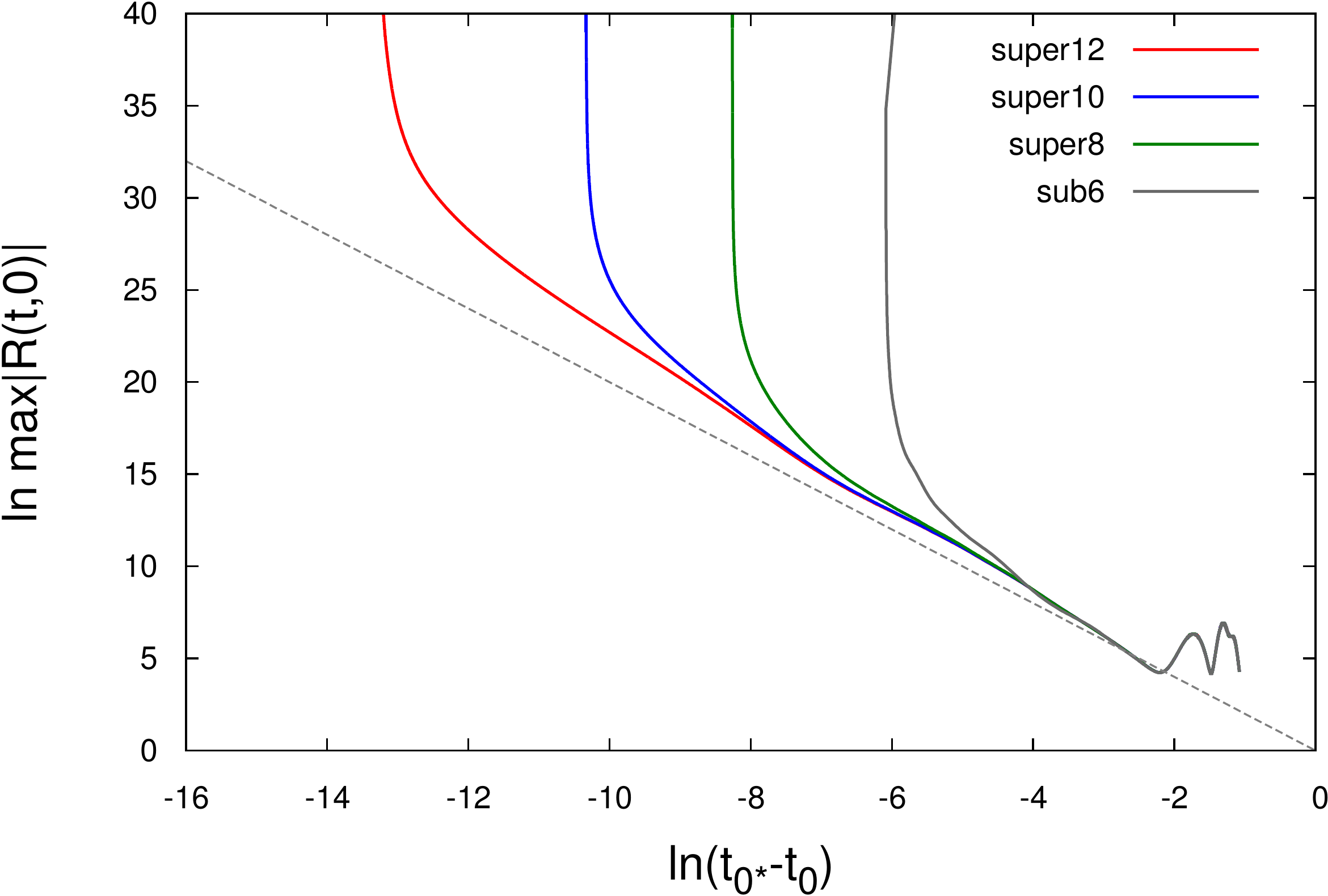}
\includegraphics[scale=0.3, angle=270]{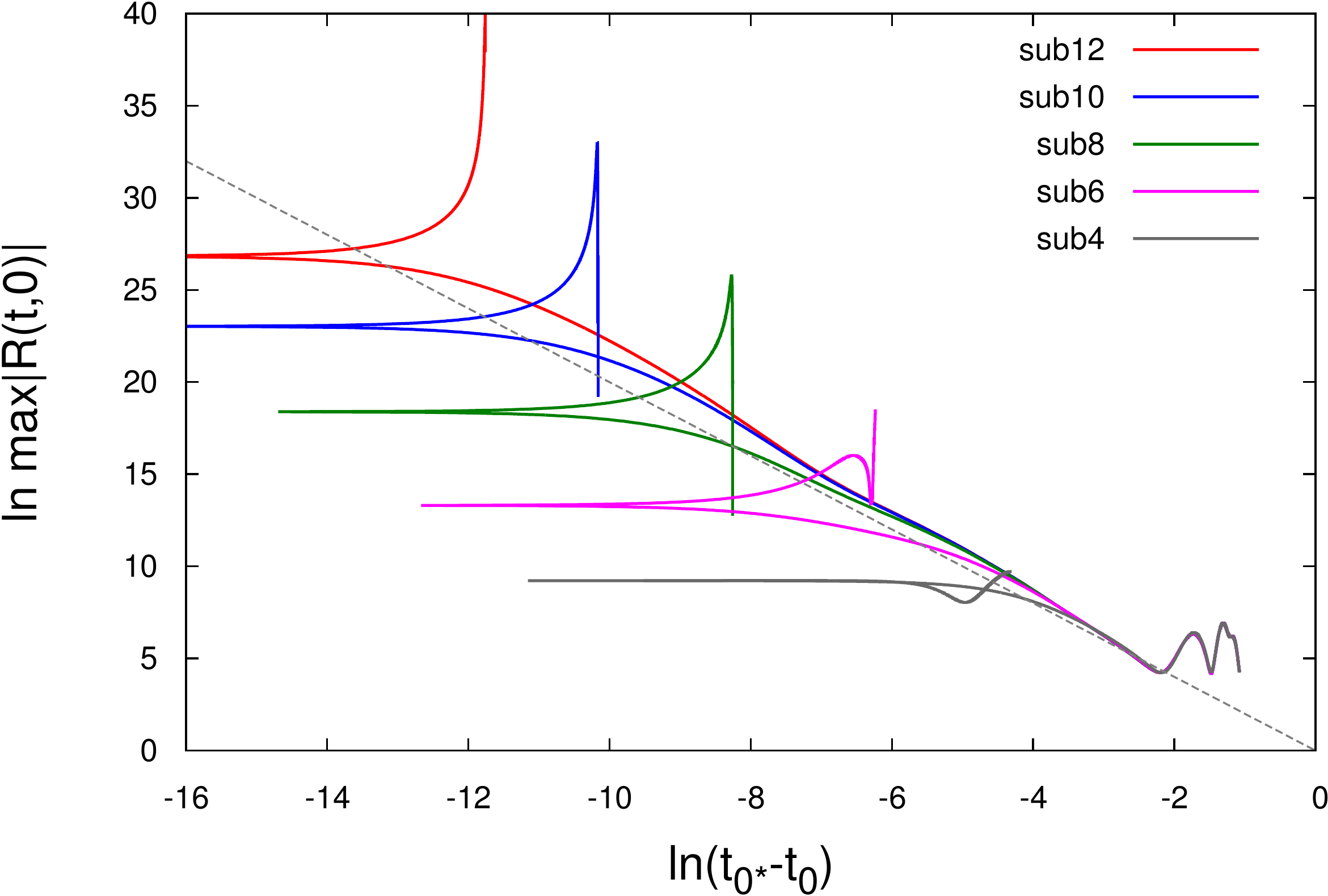}
\caption{$m=1$ D data: the equivalent plots to
  Fig.~\ref{figure:m0_Ar_ricci_t0}.}
\label{figure:m1_Dr_ricci_t0}
\end{figure}

\begin{figure}[!htb]
\centering
\includegraphics[scale=0.3, angle=270]{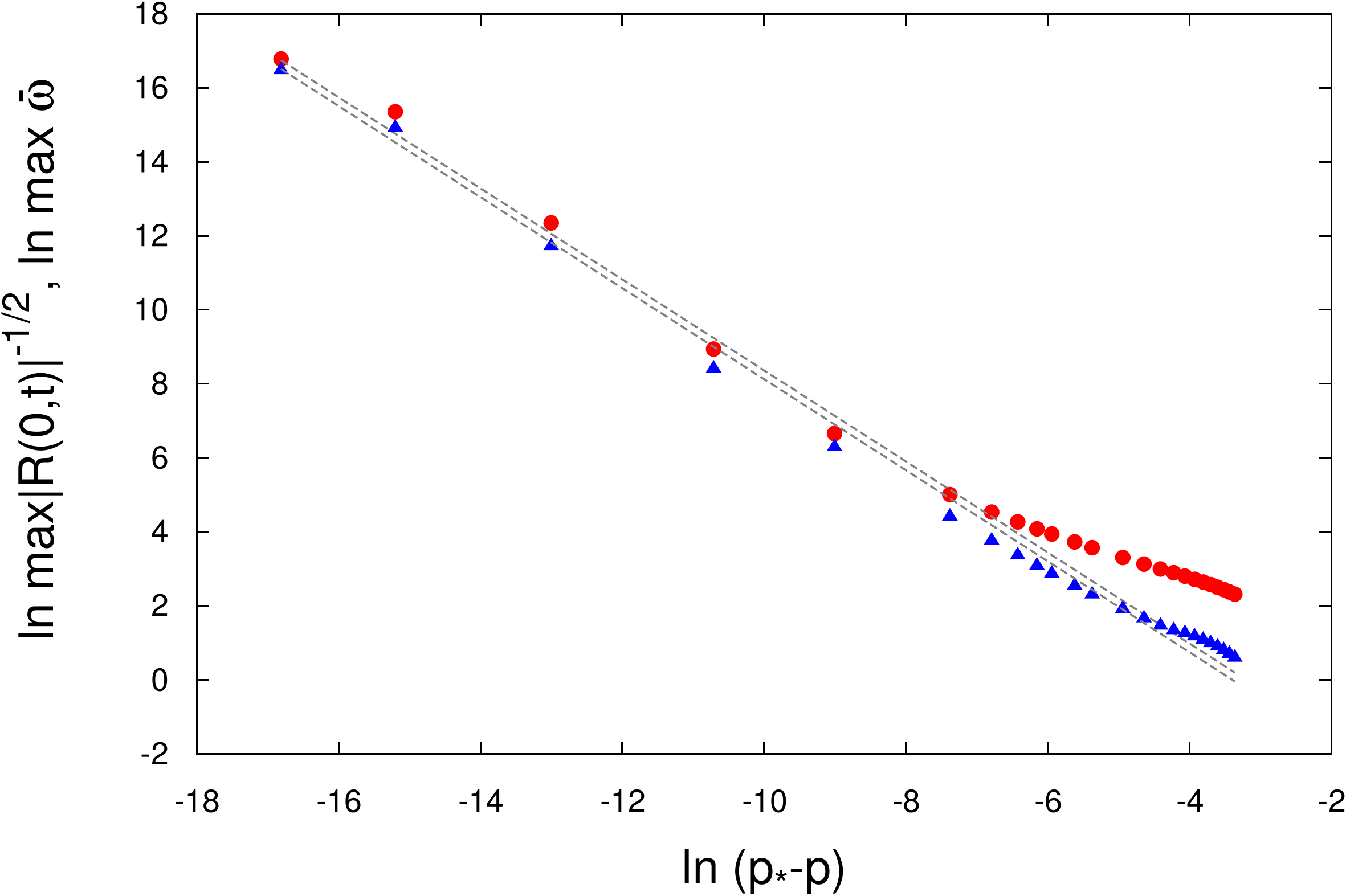}
\includegraphics[scale=0.3, angle=270]{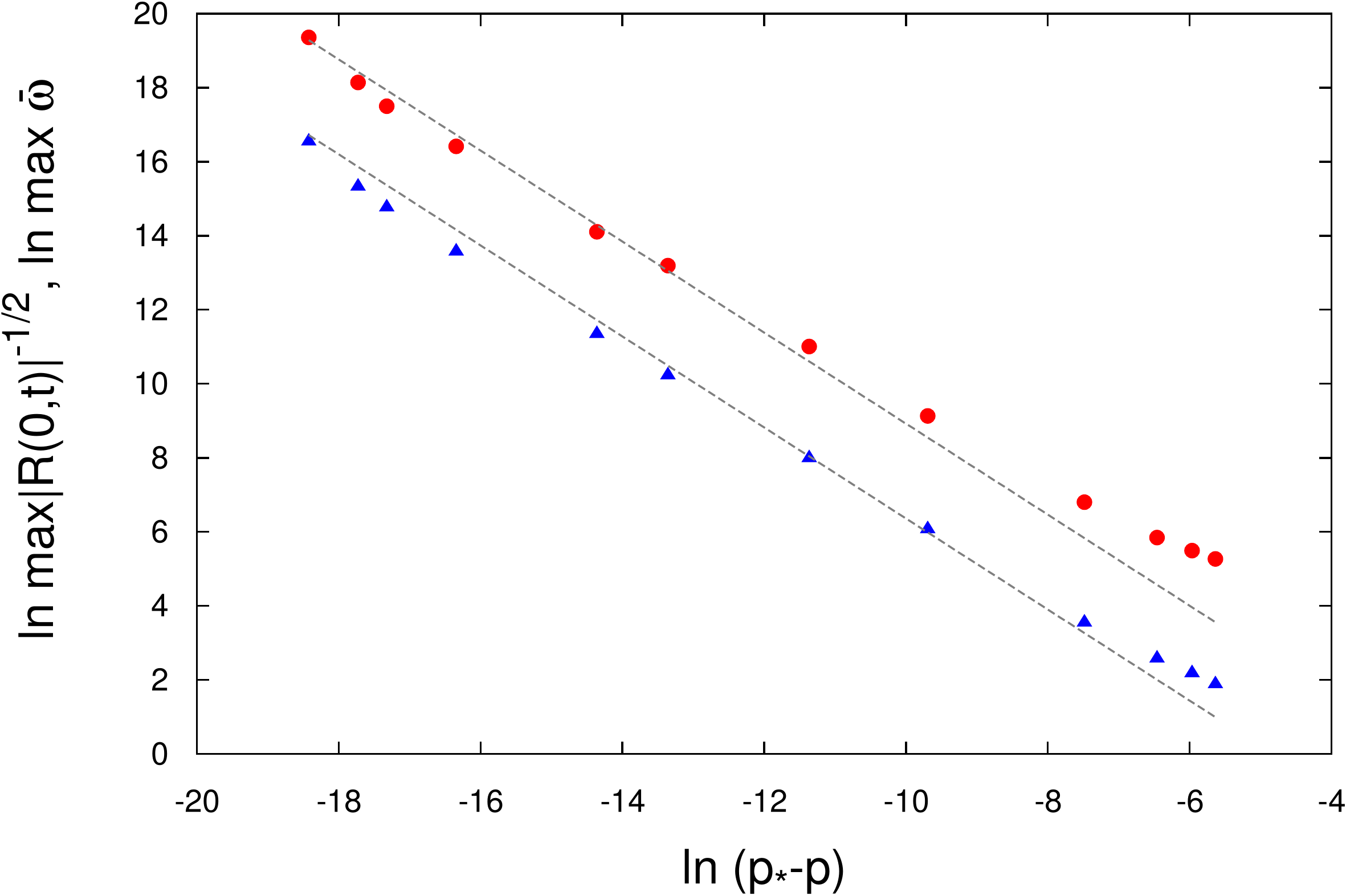}
\caption{$m=0$ B (top) and C (bottom) data: power law scaling of
  $\max_t |R(t,0)|^{-1/2}$ (full red dots) and $\max_t\omega(t,0)$,
  against $p_*-p$. These maxima occur at $\bar r =0$. The grey dotted
  lines have slope $-\gamma=-1.23$ in both plots.}
\label{figure:m0_BrCr_rot_ricci_scal}
\end{figure}

\begin{figure}[!htb]
\centering
\includegraphics[scale=0.3, angle=270]{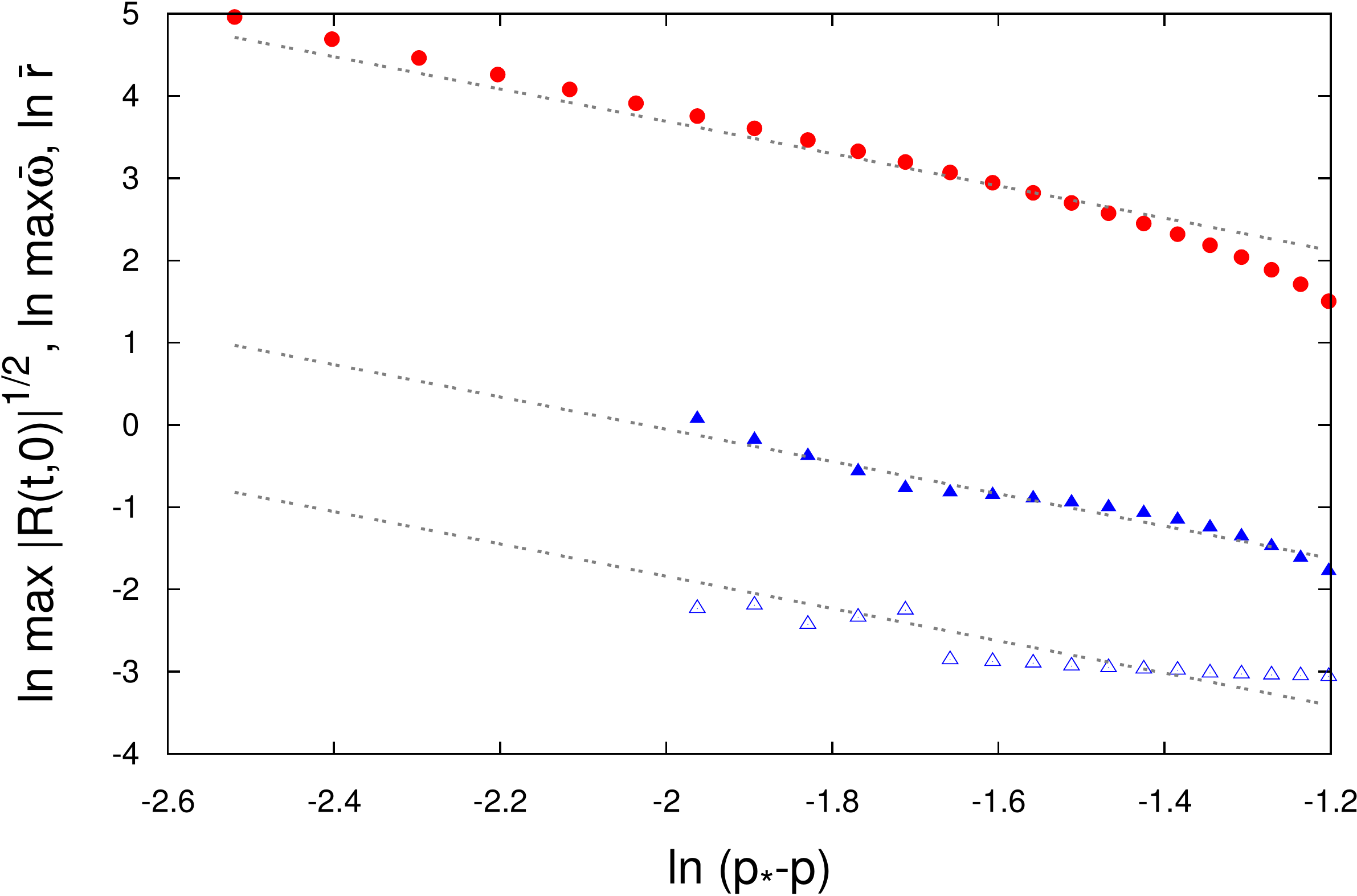}
\includegraphics[scale=0.3, angle=270]{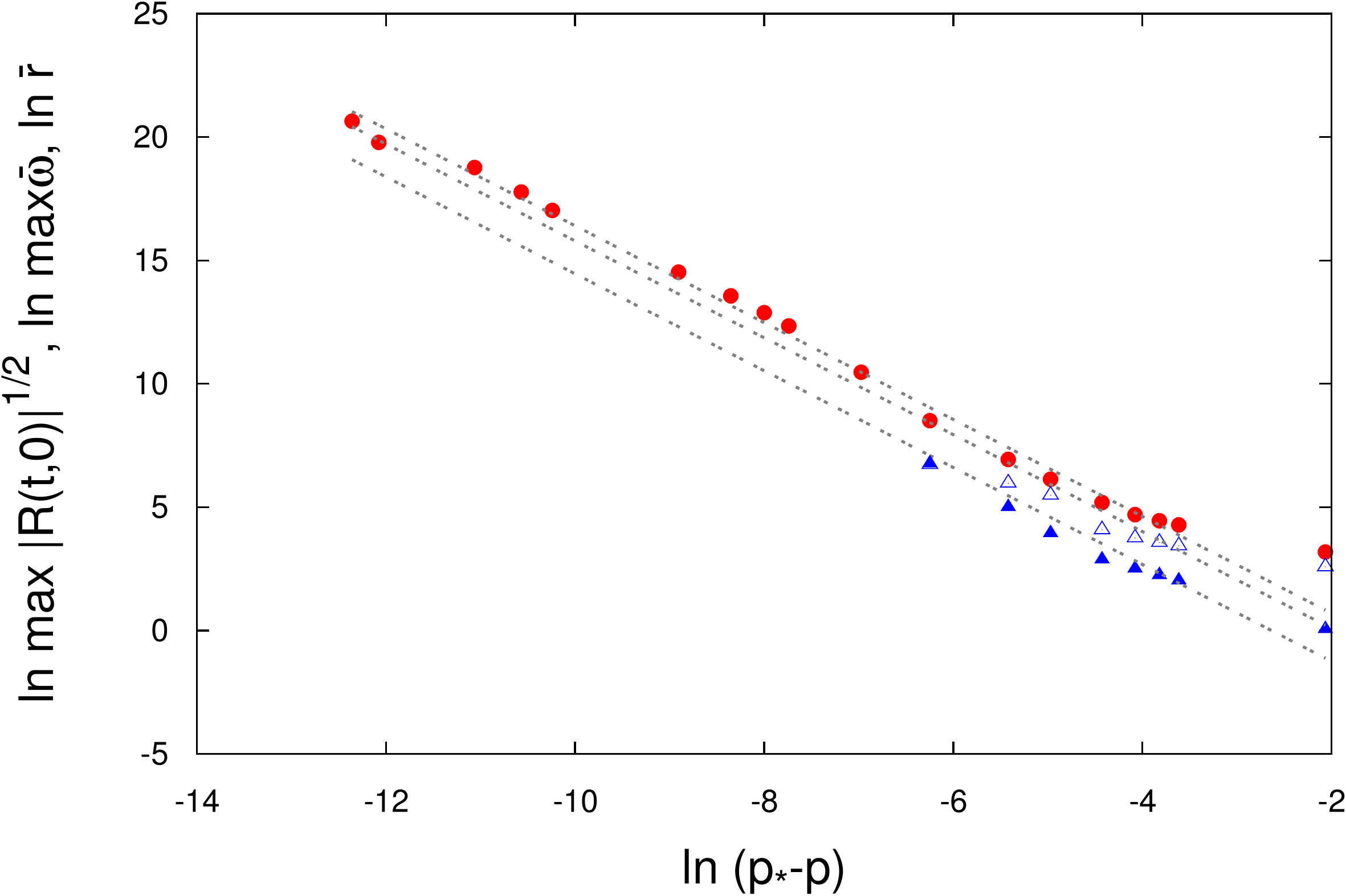}
\caption{$m=1$ B (top) and D (bottom) families: power law 
  scaling of $\max_t R(t,0)^{1/2}$ (red dots), $\max_{t,r} \bar
    \omega(t,r)$ (blue triangles) and the location in $\bar r$ of the
    maximum of $\bar\omega$  (blue empty triangles). The slope of
    all fitting lines is $-\gamma=-1.93$, with the vertical offset fitted
    individually (and clearly differing between the two families).}
\label{figure:m1_BrDr_rot_ricci_scal}
\end{figure}

\begin{figure}[!htb]
\centering
\includegraphics[scale=0.3, angle=270]{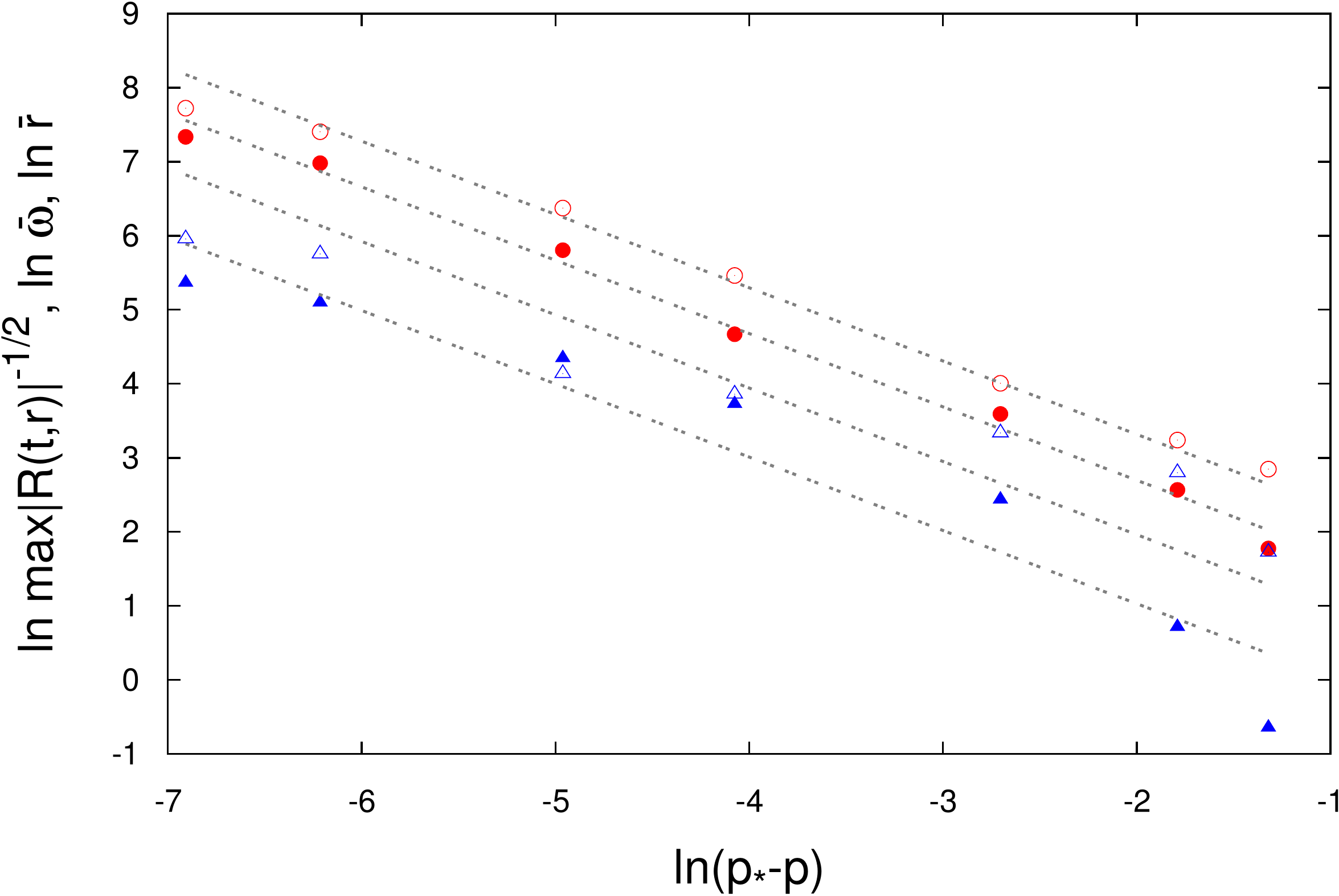}
\includegraphics[scale=0.3, angle=270]{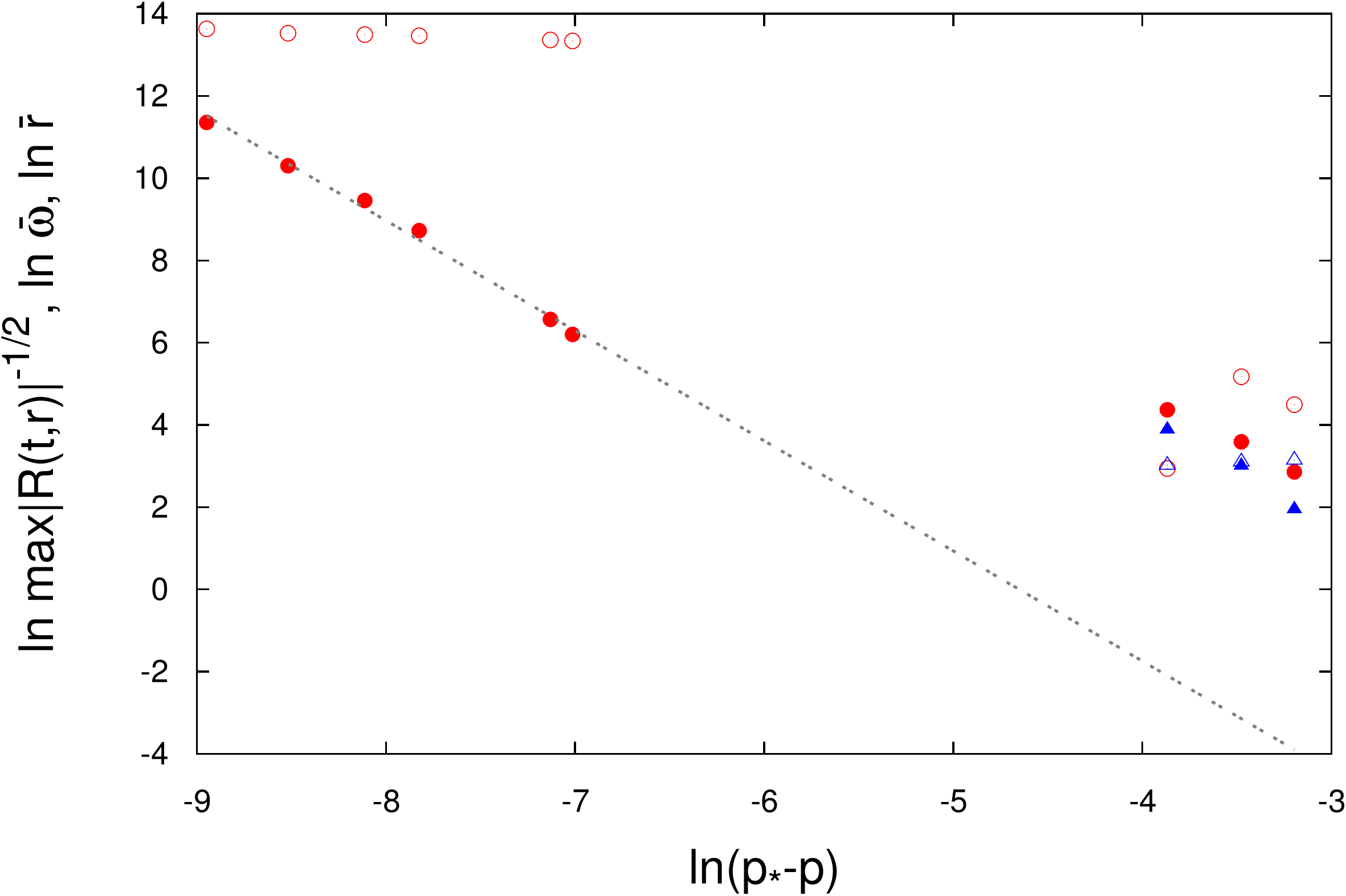}
\caption{$m=2$ B (top) and C (bottom) data: power law 
  scaling of $\max_{t,r} R(t,r)^{1/2}$ (red dots), $\max_{t,r} \bar
    \omega(t,r)$ (blue triangles) and the location in $\bar r$ of the
    maxima of $R$ (red empty dots) and $\bar\omega$  (blue empty triangles).}
\label{figure:m2_BrCr_rot_ricci_scal}
\end{figure}

Figs.~\ref{figure:m0_Ar_ricci_t0} and \ref{figure:m1_Dr_ricci_t0}
provide evidence for the behaviour (\ref{RCSS}) and (\ref{RpostCSS})
for the $m=0$ A and $m=1$ D families of initial data, by showing $\ln
R(t,0)$ against $\ln|t_0-t_{0*}|$.

In asymptotically flat spacetime, $f_-$ has a single maximum before
decaying to zero. This means that, for subcritical ($p<p_*$) data
\begin{equation}
\max_t|R(t,0)|\simeq bA^2(p_*-p)^{-2\gamma},
\end{equation}
where $A$ is again the same family-dependent constant as in
(\ref{t0nonlin},\ref{RpostCSS}) and $b:=\max f_-$ is a universal
dimensionless constant \cite{GarfinkleDuncan}. [This is a more
  explicit version of (\ref{gammadef}) above.] In our investigation
\cite{adscollapsepaper} of the $m=0$ case, we found that $f_-$ had two
maxima and two minima before the final blowup. We could demonstrate
scaling of both the values, and the location in $t_0-t_{0*}$, of all
these extrema. Similar scaling laws hold for other geometric
invariants, such as $\omega$ defined in (\ref{omegadef}), for which
self-similarity predicts
\begin{equation}
\label{omegagamma}
\omega_{\rm max}\sim (p_*-p)^{-\gamma}.
\end{equation}
For $\omega$ we only find a single maximum.

Fig.~\ref{figure:m0_BrCr_rot_ricci_scal} gives evidence of these scaling
laws for the $m=0$ B and C families,
Fig.~\ref{figure:m1_BrDr_rot_ricci_scal} for the $m=1$ B and D families,
and Fig.~\ref{figure:m2_BrCr_rot_ricci_scal} for the $m=2$ B and C
families.

For quantities which vanish identically at the centre, such as the
Ricci scalar for $m\ge 2$, or the angular momentum density $\omega$
for any $m$, we look for the maximum over all $r$ for a given $t$
instead, and plot this against $t_0-t_{0*}$. The resulting function of
$t$ clearly depends on the time slicing, but seems to scale anyway,
see again Figs.~\ref{figure:m1_BrDr_rot_ricci_scal} and
\ref{figure:m2_BrCr_rot_ricci_scal}.

Table~\ref{table:fits} shows the value of the Ricci scaling exponent
$2\gamma$ for our 12 families of initial data. For $m=0$, $\gamma$ is
the same for all families, as one would expect if there was a unique
CSS solution with a single unstable mode. Strikingly, for $m>0$,
$\gamma$ depends strongly on the family. 

Beyond looking at the behaviour of the global maxima of $R$ and
$\omega$, or the behaviour of their maxima over $r$ as a function of
$t$, we have also attempted to look for direct evidence of
self-similarity as a function of $(r,t)$, as we did successfully in
\cite{adscollapsepaper} for the $m=0$ case. We have constructed
double-null coordinates $\tilde u$ and $\tilde v$ normalised to be
proper time at the origin and with their origins fixed so that $\tilde
u=\tilde v$ at the centre and $\tilde u=\tilde v=0$ at the
accumulation point $t=t_*$. We can then define coordinates adapted to
the self-similarity as $T:=-\ln(-\tilde u)$ and $x:=\tilde v/\tilde
u$, and plot against these coordinates. Quantities such as
$\phi^2+\psi^2$, $e^{-T}\omega$, $M$, $R^{-2T}$ should then be
functions of $x$ only in any self-similar region. However, the only
quantity for which this works is the Ricci scalar. For this reason,
we do not show any plots of, for example, the scalar field.

\subsection{EMOTS location}
\label{section:emotslocation}

As already discussed above, the AH is the curve in the $tr$ plane
defined by $\rb_{,v}=0$ , so that every point on it is a MOTS. As in
\cite{adscollapsepaper}, we denote a local minimum of $t_{\rm AH}(r)$
as an earliest MOTS (EMOTS). If there are two (or more) EMOTS, then in
\cite{adscollapsepaper} we denoted the earliest of these as the first
MOTS (FMOTS), but we did not find this behaviour for $m>0$.

\begin{figure}[!htb]
\centering
\includegraphics[scale=0.45, angle=270]{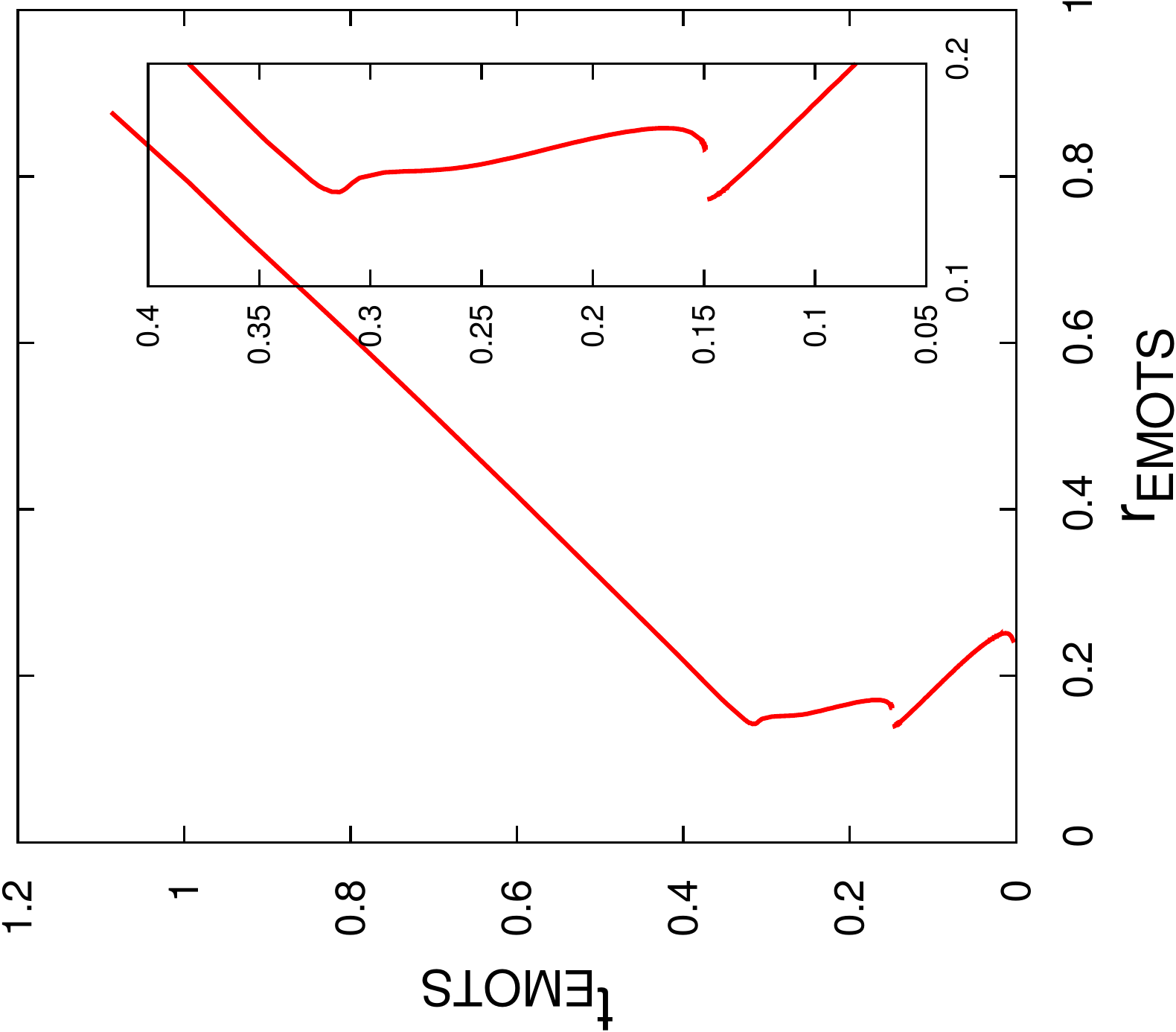}
\caption{$m=1$ A data: the trajectory of the EMOTS
  location in the $tr$ plane, for values of the scalar field amplitude
  $p$ from $0.74$ (bottom) to $0.405$ (top). The $r$ and $t$ axes are
  drawn to the same scale so that null curves are at 45 degrees. The
  inset shows the timelike segment of the curve. We believe that the
  curve is actually continuous where there appears to be a break.}
\label{figure:m1_Ar_EMOTS_tr}
\end{figure}

We focus here on the dependence of the entire AH curve, and the EMOTS
location as one aspect of this, as a function of the parameter $p$,
taking the example of the $m=1$ A data. A MOTS is already contained in
the initial data for $p\gtrsim 0.74$. Reducing $p$ from this value,
the location of the EMOTS moves inwards on an approximately null
curve, then moves outwards very rapidly in $p$ in a spacelike
direction at $p=p_{\rm break}\simeq 0.445465$, then moves to the
future on a timelike curve, a little inwards again and then outwards
again on an approximately null curve. Fig.~\ref{figure:m1_Ar_EMOTS_tr}
illustrates this for the range $0.74>p>0.404$.

\begin{figure}[!htb]
\centering
\includegraphics[scale=0.45, angle=270]{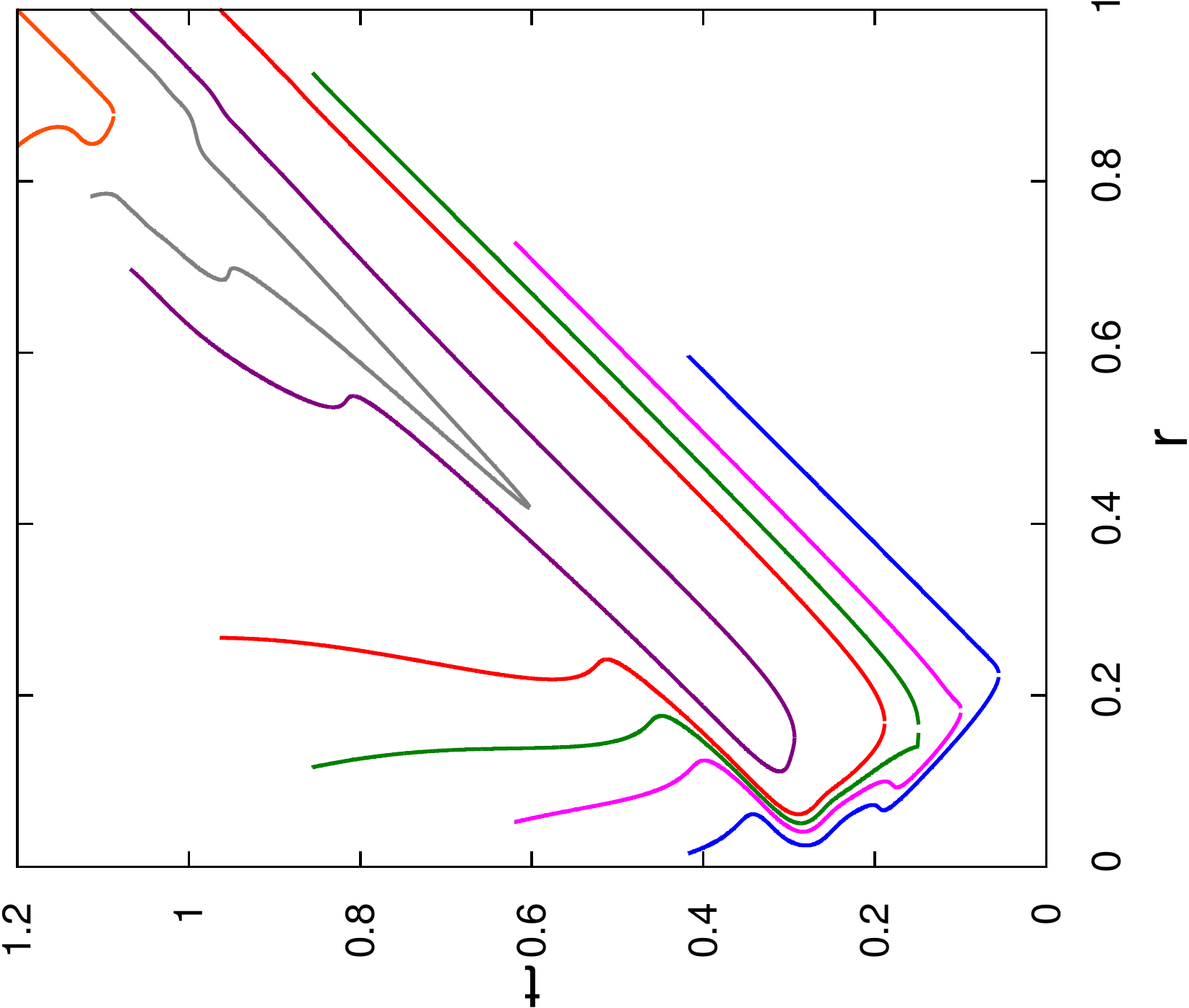}
\includegraphics[scale=0.32, angle=270]{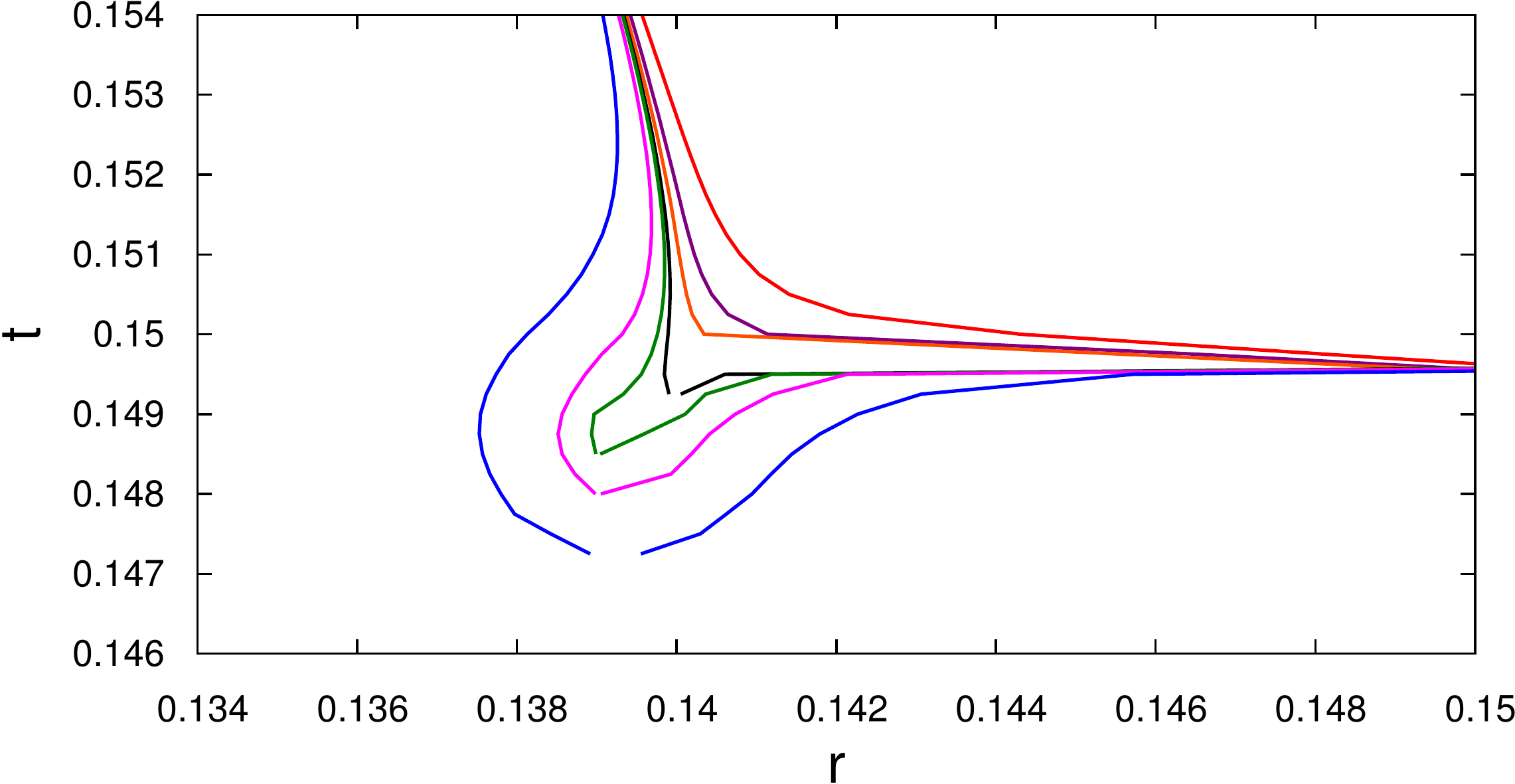}
\caption{$m=1$ A data: the top plot shows the AH for
  representative values of $p$, namely $p=0.405$,
  $0.41$, $0.415$, $0.435$, $p_{\rm break}\simeq 0.445465$, $0.46$ and
  $0.5$ (from top to bottom). The bottom plots shows values of $p$
  closer to $p_{\rm break}$, namely $p=0.445460$, $\dots 66$, $68$,
  $694$, $70$, $72$, $80$ (from top to bottom).}
\label{figure:m1_Ar_AHplots}
\end{figure}

Fig.~\ref{figure:m1_Ar_AHplots} shows how this comes about, by showing
the AH curve for selected values of $p$, with the lower plot zooming
in on $p_{\rm break}$. For all $p$, there is only a single EMOTS, but
the nature (timelike/spacelike) of the AH curve for the $m=1$ A data
varies with $p$ in a complicated way. 

For $p\simeq p_{\rm break}$ the AH curve has a section that is almost
parallel to the time slices, and the local minimum moves along that
shallow section very quickly, giving rise to the apparent jump in
Fig.~\ref{figure:m1_Ar_EMOTS_tr}. This behaviour of the EMOTS is
highly slicing-dependent.

For $p<p_{\rm break}$, the EMOTS mass does not scale. This is
reminescent of the $m=0$ A data investigated in detail in
\cite{adscollapsepaper}: in that case there were two EMOTS, with a
discontinuous switch from the inner to the outer EMOTS being the
FMOTS. Only the inner EMOTS scaled.

\begin{figure}[!htb]
\centering
\includegraphics[scale=0.45, angle=270]{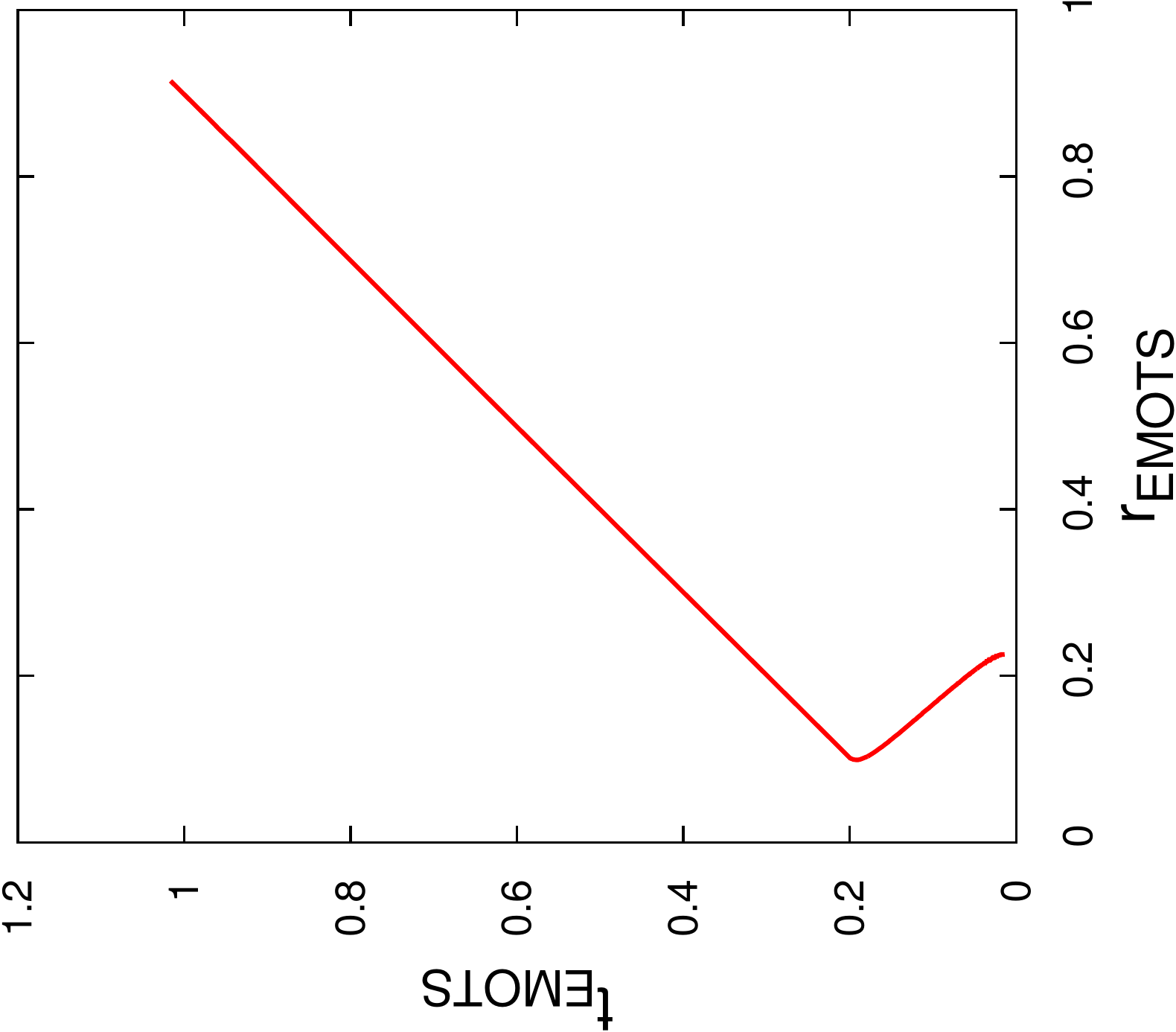}
\caption{$m=1$ C data: the trajectory of the EMOTS
  location in the $tr$ plane, for values of the scalar field amplitude
  $p$ from $0.07$ (bottom) to $0.044$ (top). The extended timelike segment is
  not present for this family of initial data.}
\label{figure:m1_Cr_EMOTS_tr}
\end{figure}

Fig.~\ref{figure:m1_Cr_EMOTS_tr} shows the EMOTS trajectory for the
$m=1$ C data. This is much simpler, and the transition from the
ingoing null to the timelike segment is now clearly continuous.

\subsection{EMOTS mass and angular momentum}
\label{section:emotsmass}

A key observation is that for $m=0$ there is a single critical value
$p_*$ governing both subcritical and supercritical scaling, while for
$m>0$ we have very different critical values of $p$ for subcritical
scaling of the maximum of the Ricci scalar, and for scaling of the
EMOTS mass, with $p_{*M}<p_*$. This means that both cannot be
controlled by the same critical solution (in contrast to the $m=0$
case, where we have a theoretical model for this that also predicts
the critical exponents).

For $m>0$, the evidence for supercritical EMOTS mass scaling is even
weaker than for subcritical $R$ and $\omega$ scaling, and for the EMOTS
angular momentum we have not found any scaling. Therefore, in
the following, we do not show log-log plots of $M$ and $J$ against
$p$, but show $p$ on a linear scale. As for the subcritical scaling of
$R$ and $\omega$, the supercritical scaling of $M_{\rm EMOTS}$ does
not continue to arbitrarily small scales for $m>0$. The mass scaling
exponents, and the ranges of $\ln(p-p_{*M})$ for which we observe
approximate power-law behaviour, are listed in
Table~\ref{table:fits}. Like $\gamma$, the mass scaling exponent
$\delta$ depends strongly on the family of initial data.

\begin{figure}
\centering
\includegraphics[scale=0.3, angle=270]{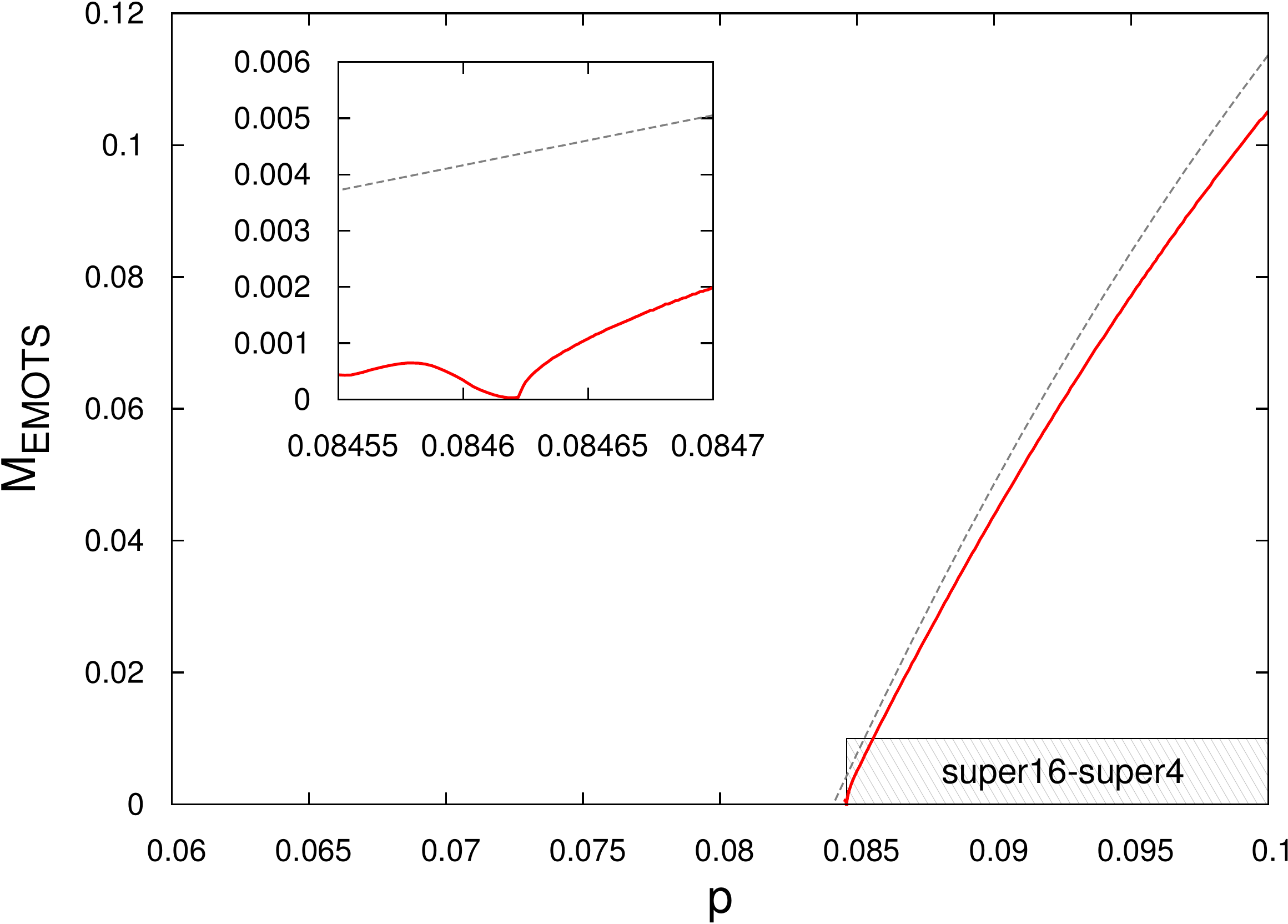}
\includegraphics[scale=0.3, angle=270]{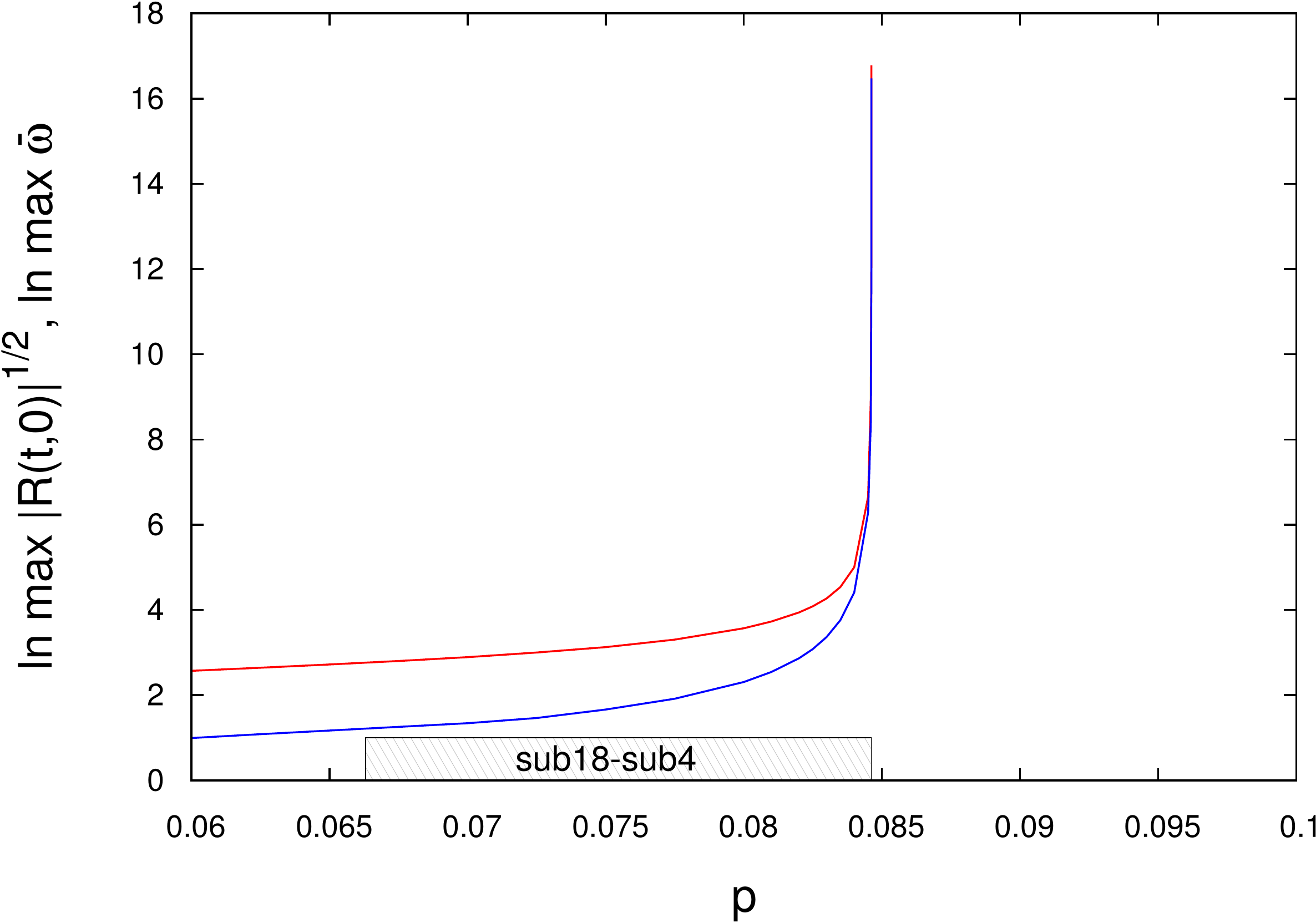}
\caption{$m=0$ B data: the upper plot shows the EMOTS mass (red) and
  total mass (grey) against $p$. The inset shows the FMOTS location
  jump due to the presence of an inner EMOTS formed later in time
  (compare with Fig.~6 of \cite{adscollapsepaper}). The lower plot
  shows $\max_tR(t,0)$ (red) and $\max_{t,r}\bar\omega(t,r)$ (blue)
  against the same range of $p$. Note that $p_{*M}=p_*$ for this and
  all other $m=0$ families only.  The shaded grey strips on the
  $p$-axis indicate the approximate ranges of $p$ where we observe
  scaling, with the text in these strips also indicating the
  corresponding range of $-\ln(p-p_{*M})$ and $-\ln(p_*-p)$, respectively.}
\label{figure:m0_Br_linlinplots}
\end{figure}

\begin{figure}
\centering
\includegraphics[scale=0.3, angle=270]{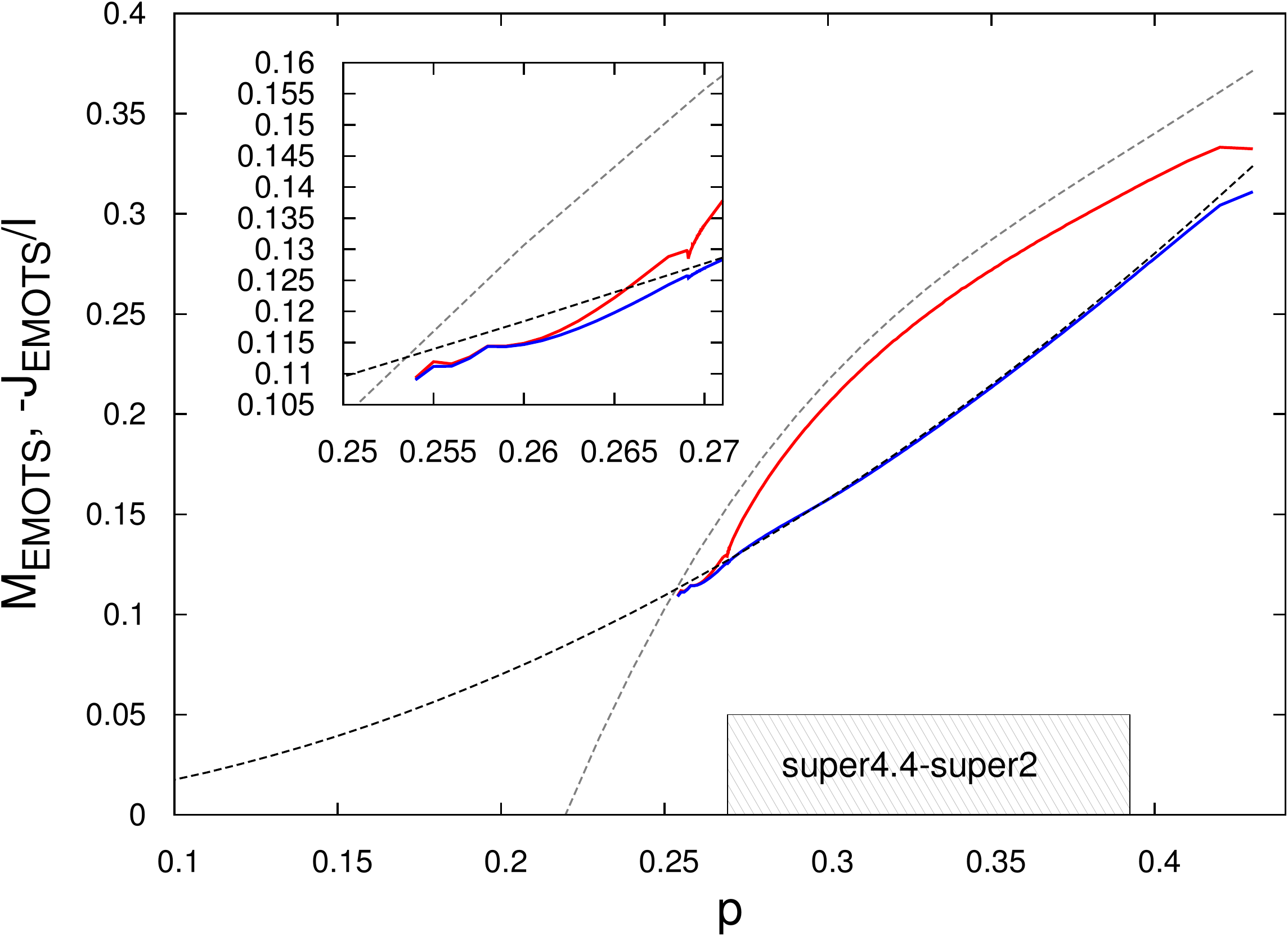}
\includegraphics[scale=0.3, angle=270]{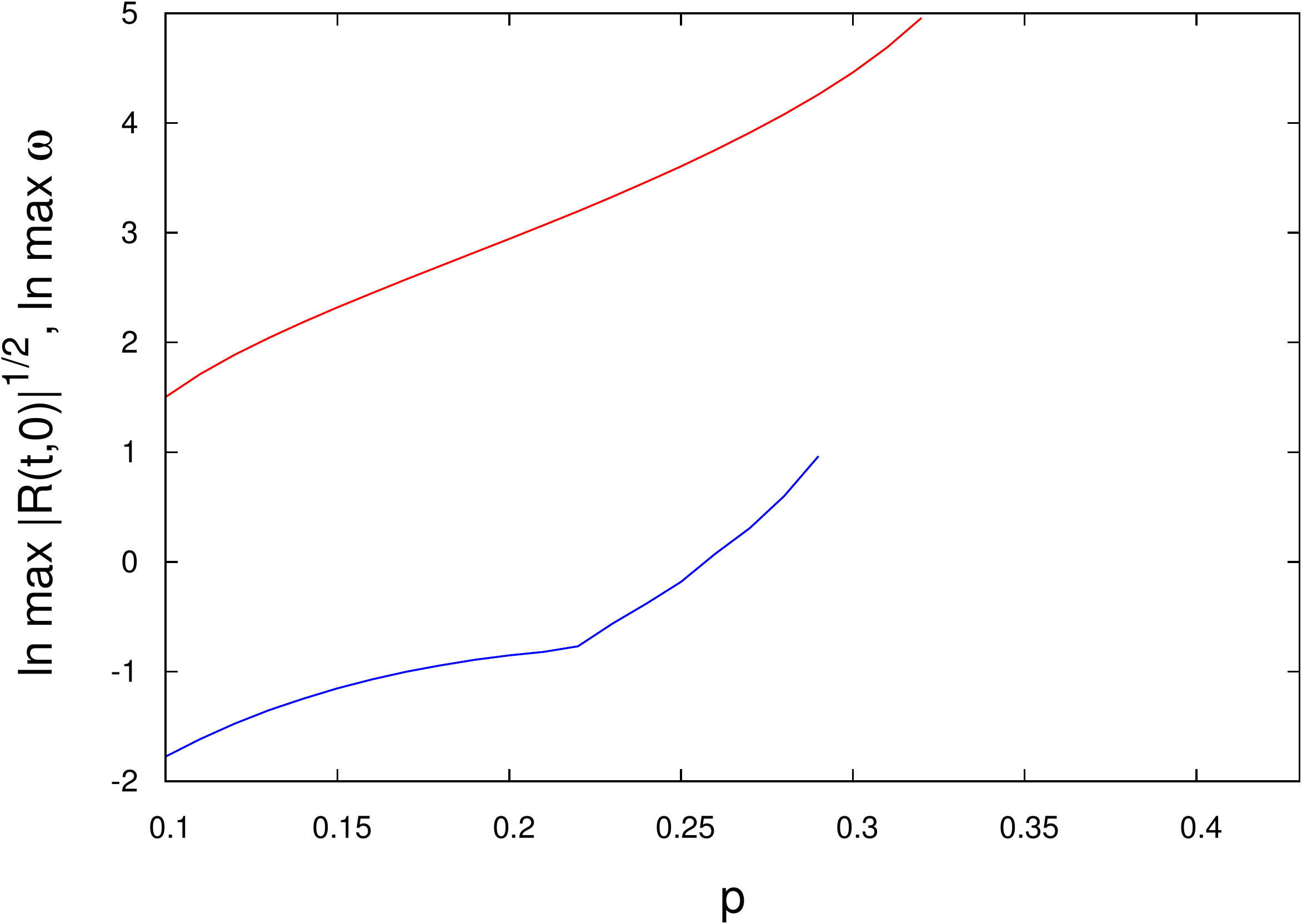}
\caption{$m=1$ B data: the equivalent plots to
  Fig.~\ref{figure:m0_Br_linlinplots}, except that now $\max\omega$ is
  shown instead of $\max\bar\omega$, and $-J_{\rm EMOTS}/\ell$ is also
  shown. The EMOTS angular momentum (blue) is almost all of the total
  angular momentum (black dotted), and not surpringly does not show
  scaling. The EMOTS mass (red) is not much below the total mass (gray
  dotted), but does show scaling. We have not found any clear
  subcritical scaling. We believe this is ``covered up'' by other
  features of the time evolution.}
\label{figure:m1_Br_linlinplots}
\end{figure}

\begin{figure}
\centering
\includegraphics[scale=0.3, angle=270]{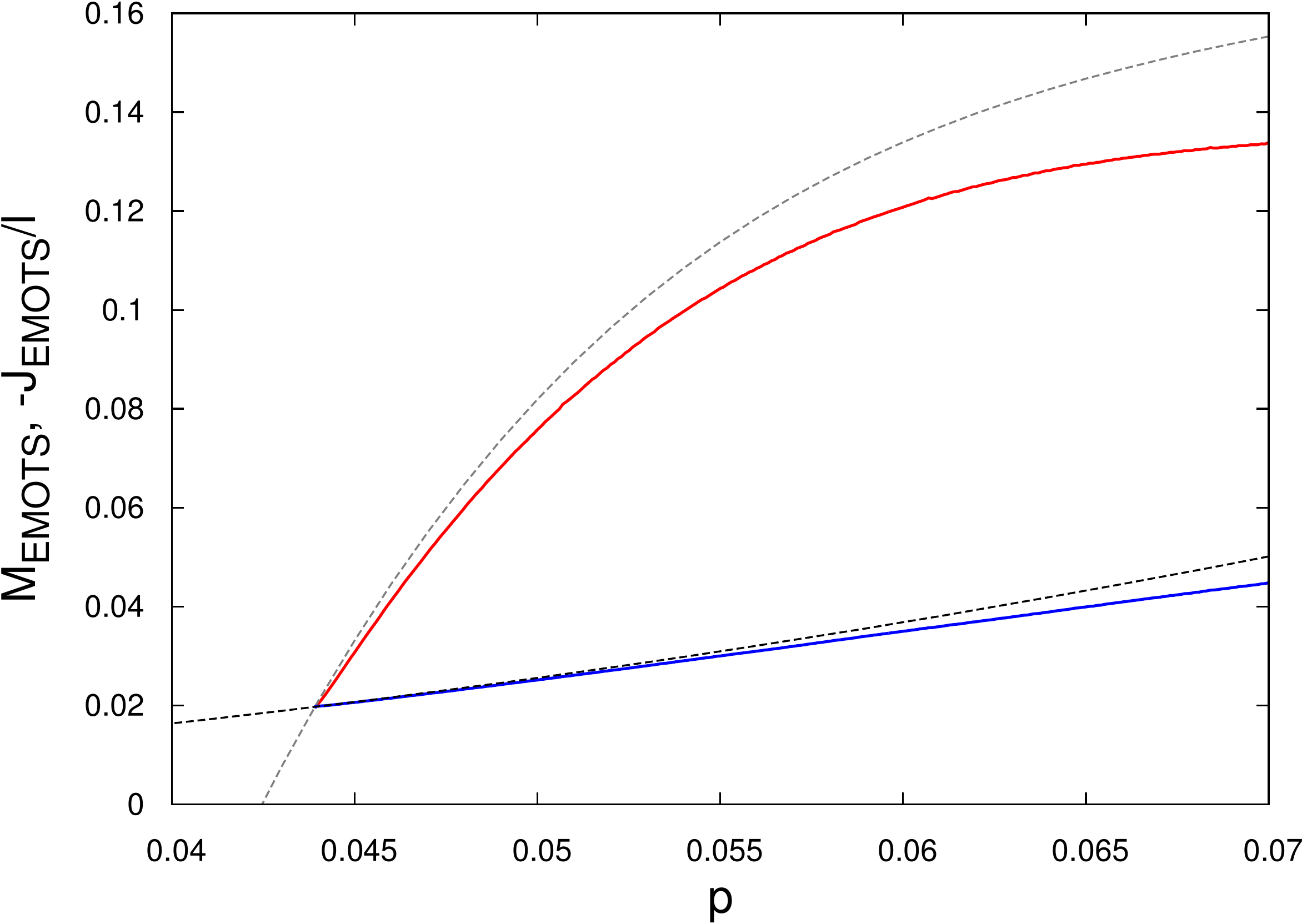}
\includegraphics[scale=0.3, angle=270]{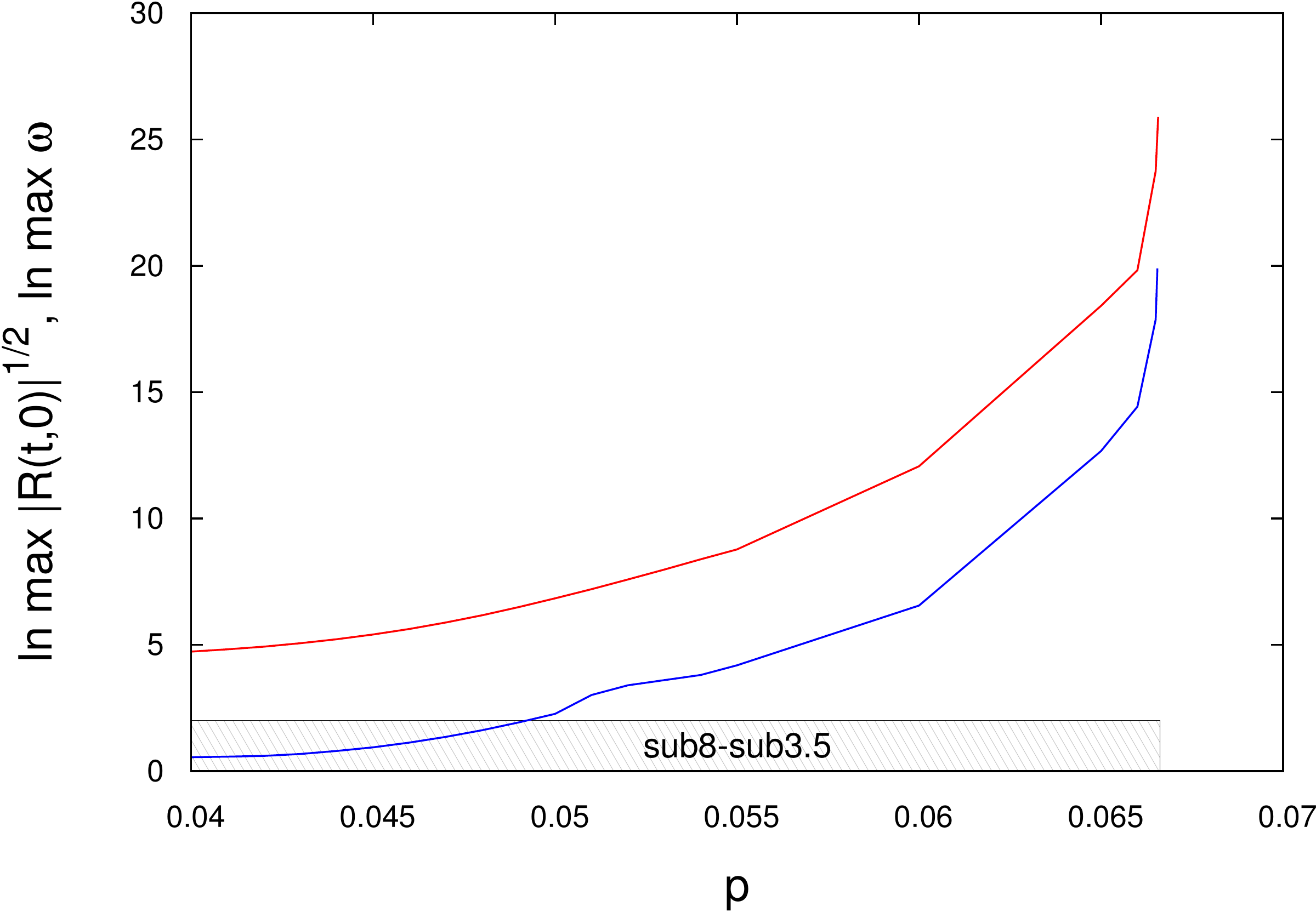}
\caption{$m=1$ C data: the equivalent plots to
  Fig.~\ref{figure:m1_Br_linlinplots}. In contrast to the $m=1$ B
  data, we find subcritical but not supercritical scaling.}
\label{figure:m1_Cr_linlinplots}
\end{figure}

\begin{figure}
\centering
\includegraphics[scale=0.3, angle=270]{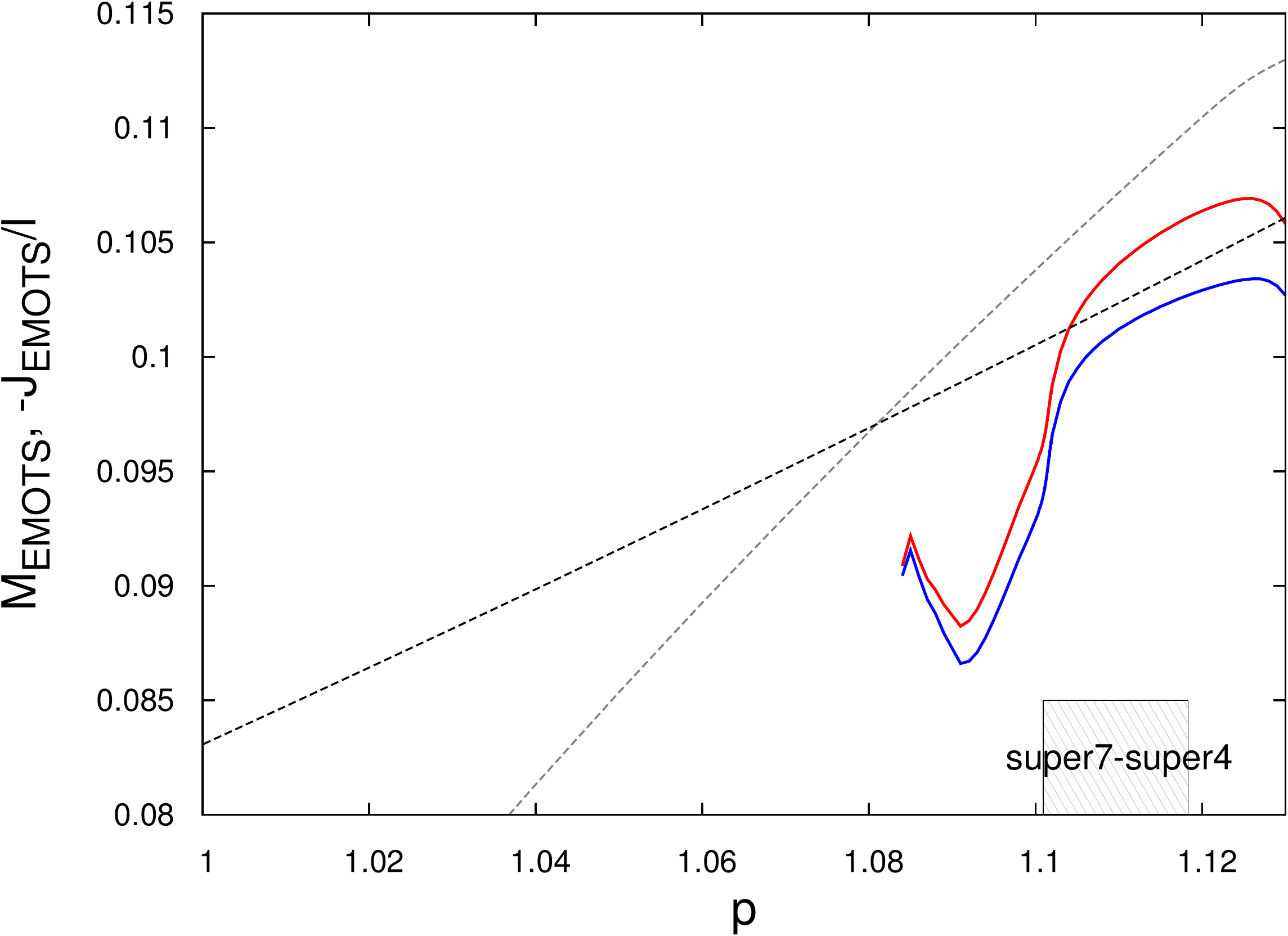}
\includegraphics[scale=0.3, angle=270]{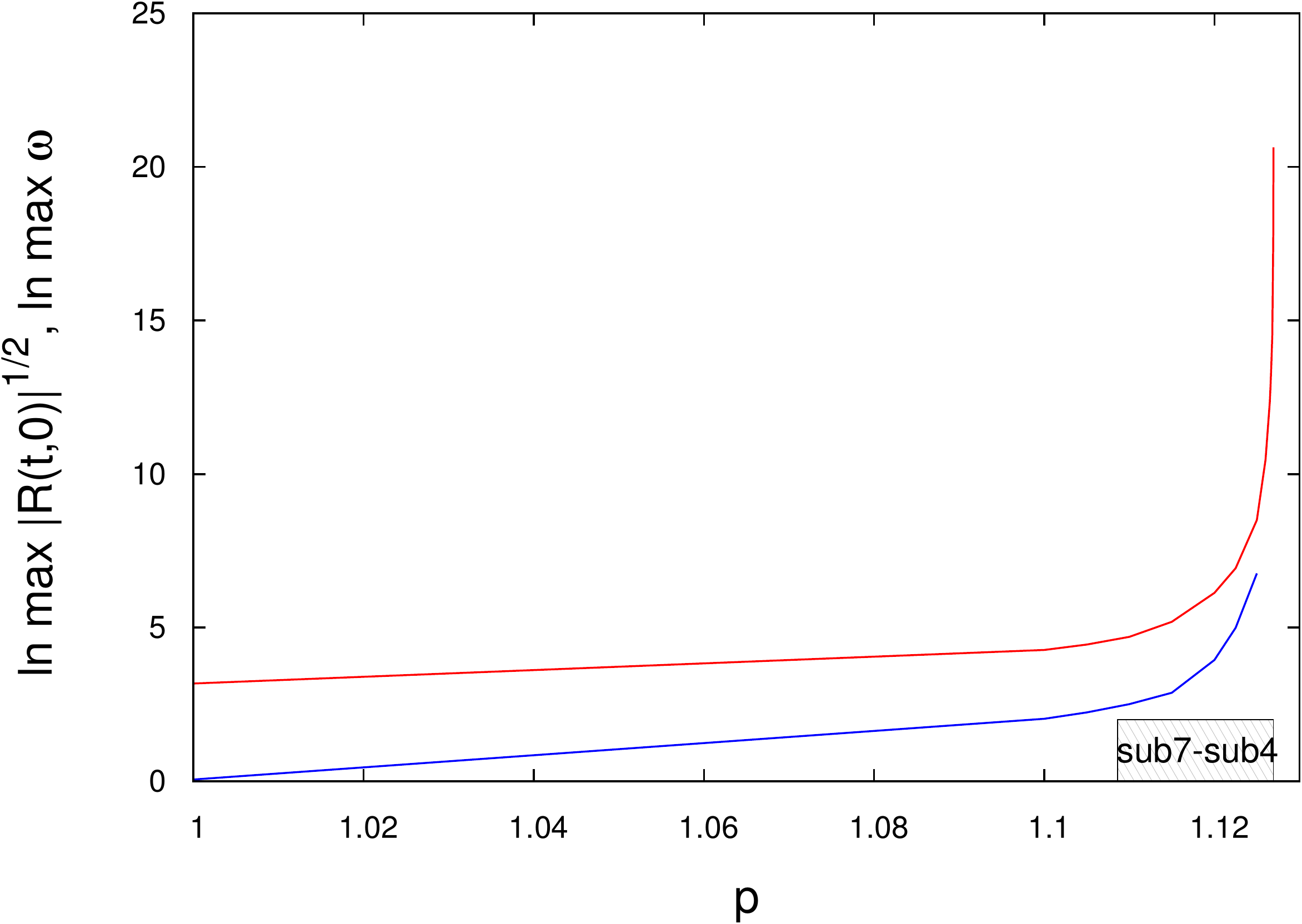}
\caption{ $m=1$ D family: the equivalent plots to
  Fig.~\ref{figure:m1_Br_linlinplots}. This is the family where we see
  subcritical scaling over the largest range of $\ln(p_*-p)$ (see the lower plot of Fig.~\ref{figure:m1_BrDr_rot_ricci_scal}).}
\label{figure:m1_Dr_linlinplots}
\end{figure}

\begin{figure}[!htb]
\centering
\includegraphics[scale=0.3, angle=270]{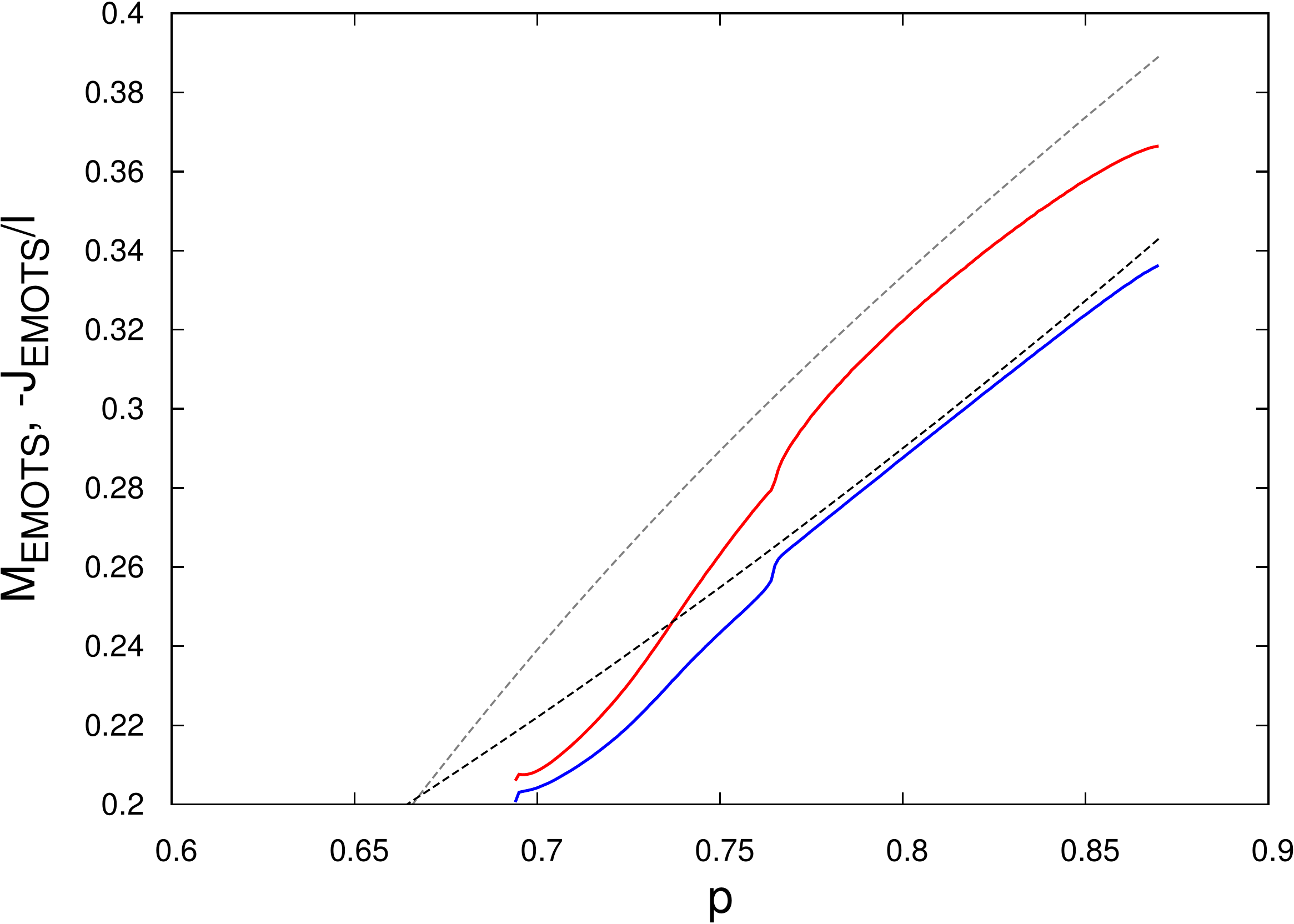}
\includegraphics[scale=0.3, angle=270]{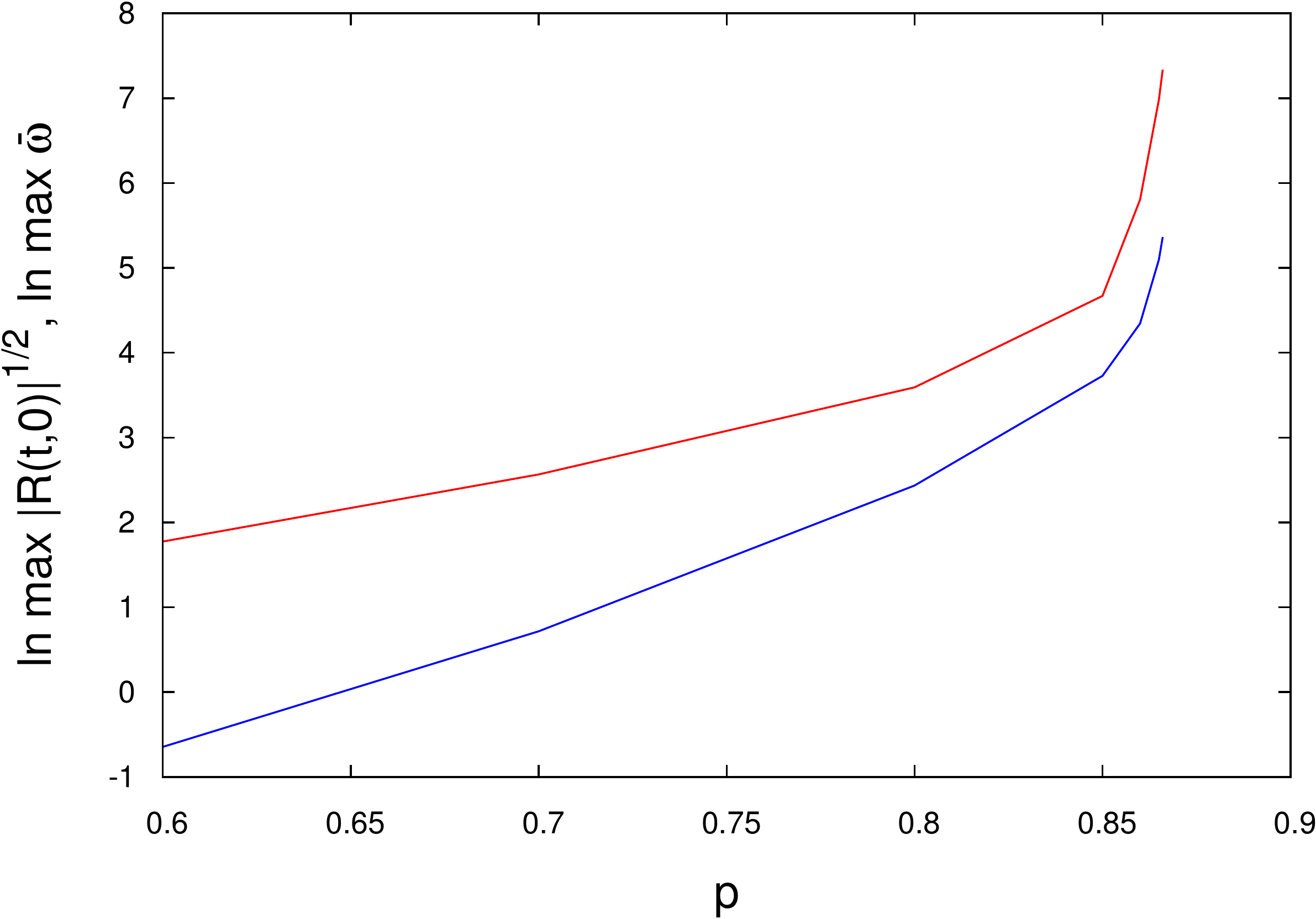}
\caption{$m=2$ B family: $M_{\rm EMOTS}$, $J_{\rm EMOTS}$, $\ln R$ and
  $\ln \omega$ versus $p$.}
\label{figure:m2_Br_linlinplots}
\end{figure}

In Figs.~\ref{figure:m0_Br_linlinplots}-\ref{figure:m2_Br_linlinplots}
we give a few examples of the behaviour of the EMOTS mass and angular
momentum, and the maxima of $R$ and $\omega$, as functions of the
amplitude $p$, over a large range of $p$. We also indicate the
approximate ranges of $p$ where we see supercritical and subcritical
scaling. In these plots, the upper end of the plotting range for $p$
corresponds to a MOTS being present already in the initial data. The
mininum of $p$ on the $M_{\rm EMOTS}$ curve corresponds to the EMOTS
location having gone back out almost to outer boundary at $t\sim 1$
(compare Figs.~\ref{figure:m1_Ar_EMOTS_tr} and
\ref{figure:m1_Cr_EMOTS_tr}), which however in $p$ is very close to
$p_{*M}$. The lower end of the plotting range for $p$ corresponds to
the lowest value of $p$ where we can clearly see a maximum of $R$ for
some $t<2$. (Finding the maximum becomes numerically very difficult,
and so it is not clear for all $p$ if one exists.)

As a reminder of the behaviour we found for the $m=0$ case in
\cite{adscollapsepaper}, Fig.~\ref{figure:m0_Br_linlinplots} shows
this for the $m=0$ B data. (For $m=0$ there is no angular momentum,
but these initial data are complex, so instead of $\omega$ we show the
``charge density'' $\bar\omega$.) This illustrates that for $m=0$
there is a single value $p_*$ controlling both supercritical and
subcritical scaling.

Figs.~\ref{figure:m1_Br_linlinplots}-\ref{figure:m1_Dr_linlinplots}
then show three different families of initial data for $m=1$. The
obvious difference to $m=0$ is that we now have separate critical
values $p_*$ for subcritical scaling and $p_{*M}$ for supercritical
scaling, with $p_*>p_{*M}$ for all $m>0$ data we have
investigated. Moreover, the blowup of $\max R$ and $\max\omega$ at
$p=p_*$ is immediately obvious (and power-law scaling is then
confirmed by log-log plots such as
Figs.~\ref{figure:m0_BrCr_rot_ricci_scal}-\ref{figure:m2_BrCr_rot_ricci_scal}),
whereas the mass scaling is much less clear both by eye and in log-log
plots.

Finally, Fig.~\ref{figure:m2_Br_linlinplots} shows an example of an
$m=2$ family of initial data, namely the $m=2$ B data. 

An additional key difference between $m=0$ and $m>0$ is that for
$m=0$, both super and subcritical scaling continues down to very small
scales: the lower cutoff is either the (small) length scale set by the
cosmological constant, or appears to be a lack of numerical
resolution. In contrast, for $m>0$ scaling seems to end at some
smallish scale for dynamical reasons that we do not yet
understand. Looking at Table~\ref{table:fits}, we see that we observe
EMOTS mass scaling only over about 3 $e$-foldings in $|p-p_*|$ for all
$m>0$ data, in contrast to up to 10 $e$-foldings for $m=0$. We see
Ricci scaling for up to 20 $e$-foldings in $|p-p_*|$ for $m=0$, but
the ``best'' we have found for $m>0$ is 9 $e$-foldings. We have no
real explanation for this failure of scaling at small scales, and can
only guess that it is covered up by the infall of matter into the
self-similar region of spacetime.

Among our three examples of $m>0$ families of data, we have selected
$m=1$ D because it shows the clearest subcritical scaling (see
Table~\ref{table:fits}, and the lower part of
Fig.~\ref{figure:m1_BrDr_rot_ricci_scal}). By contrast, $m=1$ B
(Fig.\ref{figure:m1_Br_linlinplots}) shows no subcritical scaling at
all, while $m=1$ C (Fig.\ref{figure:m1_Cr_linlinplots}) shows no
supercritical scaling. We can only guess that these are extreme
examples of infalling matter covering up what would otherwise be a
self-similar region.

There appears to be no supercritical scaling of $J$ at all for any of
our ($m>0$) families. Again we have no explanation for this. Note that
the mass that scales (and which we are using in all our plots) is the
generalised Hawking mass $M_{\rm H}$ (based on the area radius, and
becoming the irreducible mass of an isolated horizon or black hole),
not the generalised BTZ mass $M_{\rm BTZloc}$ (which includes angular
momentum, and becomes the BTZ mass of a black hole).

\subsection{Numerical error}

\begin{figure}
\centering
\includegraphics[scale=0.3, angle=270]{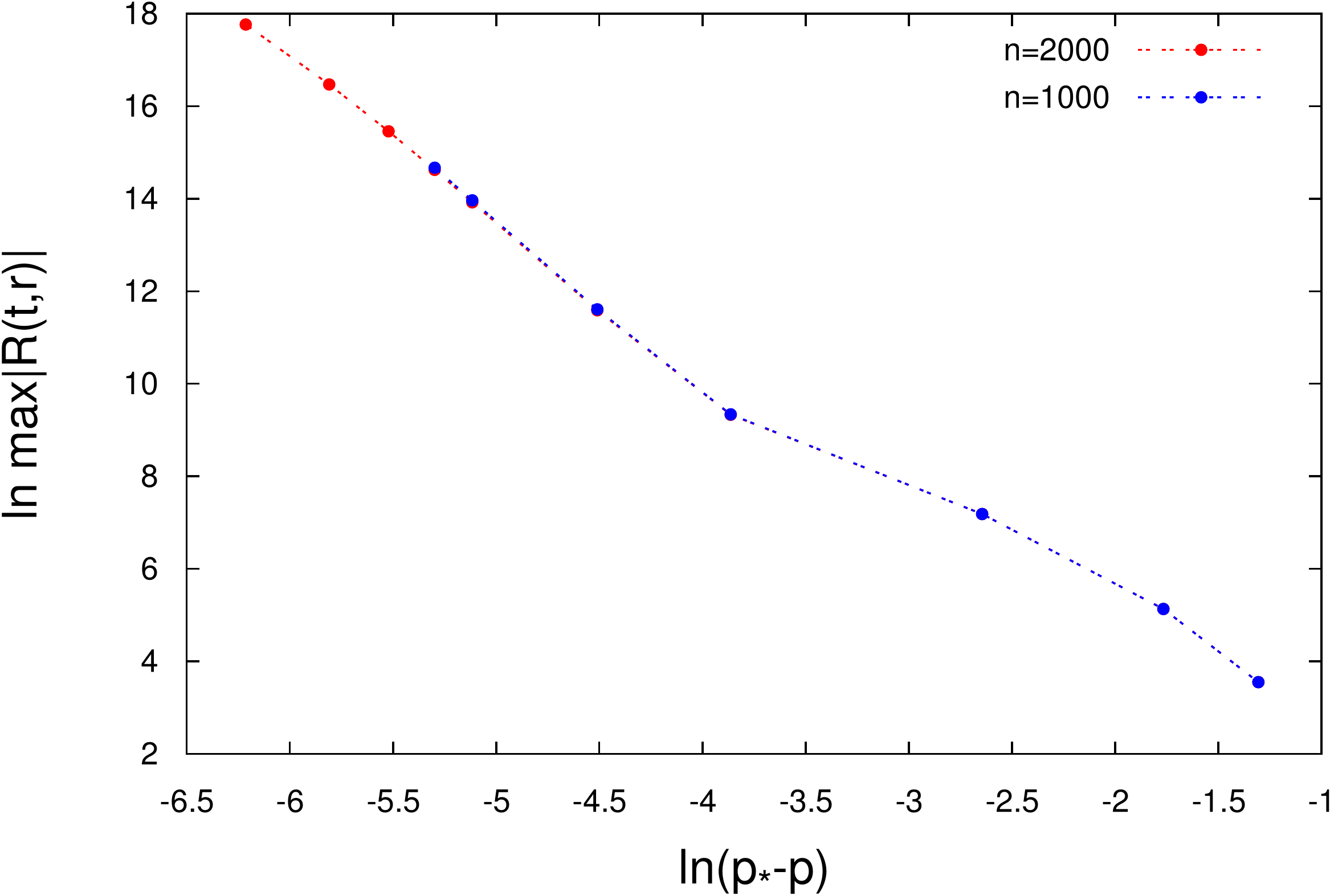}
\caption{Ricci scaling for the $m=2$ B data, as in
  Fig.~\ref{figure:m2_BrCr_rot_ricci_scal}, but with $\Delta r=1/1000$
  (blue) and $\Delta r=1/2000$ (red). A best fit by eye gives
  $p_*=0.871\pm 0.001$ at the higher resolution, and we have used
  $p_*=0.871$ at both resolutions.}
\label{figure:error}
\end{figure}

As an indication that the resolution-dependent numerical error is
small, Fig.~\ref{figure:error} shows approximate power-law scaling of
the maximum of the Ricci scalar against $p_*-p$ for 1000 and 2000 grid
points in $r$, with $\Delta r/\Delta t=1/4$ at both resolutions, for
the $m=2$ B data (see below for what these data are). Adjusting $p_*$
so that the log-log plot approaches a straight line as much as
possible, we find $p_*=0.871\pm 0.001$ at both resolutions, with
no clear difference between the value at the two resolutions. 

The main systematical error that we are aware of is a failure of our
time evolution scheme to maintain the regularity condition $A=B$ at
the centre $r=0$, at times shortly before the blowup. (In 2+1, blowup
occurs very soon after maximum curvature, as measured by the time
coordinate $t$, even for subcritical data). However, neither the EMOTS
nor the maxima of $R$ and $\omega$ occur in the domain of dependence
of the constraint violation, and so we believe their values are not
affected. 

\section{Discussion}
\label{section:discussion}

Going from the spherically symmetric scalar field in 3+1 dimensions
with $\Lambda=0$
\cite{Choptuik1993,GundlachLRR}, via the spherically symmetric scalar
field in 2+1 dimensions with $\Lambda<0$
\cite{PretoriusChoptuik,adscollapsepaper}, to the rotating
axisymmetric scalar field in 2+1 dimensions with $\Lambda<0$ (this
work) the results of numerical time evolutions become more complicated
and less well understood. Hence we begin this discussion by reviewing
the two simpler situations.

In the 3+1 case with $\Lambda=0$ 
there is a clearly defined collapse threshold $p=p_*$:
essentially all scalar field matter that does not immediately go into
making the black hole escapes to infinity instead. For arbitrary
1-parameter families of initial data, with sufficient fine-tuning one
can make the curvature arbitrarily large as $p\nearrow p_*$, and the
black hole mass arbitrarily small as $p\searrow p_*$.

In the 2+1 case we need $\Lambda<0$ to form a black hole from regular
initial data at all. This means that we effectively have a reflecting
timelike outer boundary. As a consequence, all matter eventually falls
into the black hole. In 2+1 dimensions with $\Lambda<0$ there is also
a gap in the (dimensionless) mass between the adS ground state with
$M=-1$, and the black hole solutions with $M>0$. [The range $-1<M\le 0$
  corresponds to point particles (conical singularities), which cannot
  form in collapse.]

For both these reasons the collapse threshold in 2+1 dimensions is
less clearly defined than in higher dimensions: the final black hole
mass is just the total mass $M_{\rm tot}$, as long as $M_{\rm tot}>0$,
while for $M_{\rm tot}<0$ a black hole cannot form. (By contrast, in
higher dimensions with $\Lambda<0$ there is still a reflecting
boundary condition, but now the mass has dimension, there is no mass
gap, and there is a well-defined threshold of black-hole formation
after $0,1,2,\dots$ reflections at the outer boundary \cite{BizonRostworowski}.)

In response to these features of 2+1 dimensions, we have adopted the
approach of \cite{PretoriusChoptuik}: we define as subcritical any
evolution where the Ricci scalar reaches a local maximum (at the
centre) before blowing up. We also do not measure the final black hole
mass but the mass of the first intersection of the apparent horizon
with our time slicing (the ``earliest marginally outer-trapped
surface'', or EMOTS).

In \cite{adscollapsepaper} we found empirically that for $p\simeq p_*$
the time it takes for a light ray to reach the outer boundary and come
back ($\Delta t\sim 2$ in our choice of coordinates) corresponds to an
exponentially small proper time at the centre. This means that any
outgoing radiation is scattered back to the centre almost immediately,
in terms of the relevant time at the centre, and is probably why even
in what we define as subcritical evolutions, a spacelike central
curvature singularity develops very soon after the maximum of the
Ricci scalar.

The system investigated in \cite{PretoriusChoptuik,adscollapsepaper}
is the case $m=0$ with real $\Phi$ of this work, and the particular initial data used
to produce all plots in \cite{adscollapsepaper} corresponds to the A
family of initial data here. For $m=0$, we found a critical value
$p_*$ of $p$ at which both the maximum of Ricci becomes arbitrarily
large as $p\nearrow p_*$, and the (inner) EMOTS mass arbitrarily small
as $p\searrow p_*$. We identified a continuously self-similar (CSS)
critical solution both theoretically and numerically. Based on this,
we derived the Ricci scaling exponent $\gamma$ and mass scaling
exponent $\delta$, in agreement with our numerical time evolutions.

There are some similarities between $m=0$ and $m>0$:

\begin{enumerate}

\item For most families of initial data, there is a threshold $p_{*M}$ such
that the EMOTS mass shows power-law scaling as $p\searrow p_*$. 

\item For most families of initial data, there is a threshold $p_*$ such
that the Ricci scalar shows power-law divergence as $p\nearrow p_*$. 
\label{item:Ricciscaling}

\item For initial data with $p\simeq p_*$, the maximum
of the Ricci scalar evolves as a function of proper time at centre
$t_0$ in a way that is compatible with the existence of a CSS solution
with one unstable mode -- the ``standard'' scenario for type-II
critical collapse \cite{GundlachLRR}.

\setcounter{remembercounter}{\value{enumi}}
\end{enumerate}

However, key aspects of $m>0$ also differ from $m=0$ (see 
Table~\ref{table:fits}):

\begin{enumerate}
\setcounter{enumi}{\value{remembercounter}}

\item The critical values of $p$ for Ricci scaling
and EMOTS scaling are widely separated (with $p_*>p_{*M}$ in all cases).

\item We observe both mass scaling and Ricci scaling only over a limited
range of scales. 

\item The scaling exponents $\gamma$ and $\delta$ depend strongly on the
family of initial data. \label{item:gammavaries}

\item With the exception of the Ricci scalar $R$, we have not
convincingly been able to identify a CSS or DSS spacetime-dependence
of relevant scalars such as $M$, $\phi^2+\psi^2$ or $\omega$ in
evolutions for $p\simeq p_*$. \label{item:noCSS}

\setcounter{remembercounter}{\value{enumi}}
\end{enumerate}

A strong motivation for this work was that in 2+1 spacetime dimensions
axisymmetry (with rotation) is numerically almost as straightforward
as spherical symmetry, and that the threshold of gravitational
collapse with angular momentum has hardly been studied yet. We have
made the following observations concerning angular momentum:

\begin{enumerate}
\setcounter{enumi}{\value{remembercounter}}

\item At least for some families, the maximum of the local angular
momentum density $\omega$ shows subcritical scaling with the same
(family-dependent) $\gamma$ as the maximum of $R$. \label{item:omega1}

\item However, where this is the case the constant ratio $\max R/\max
  \omega^2$ depends on the
  family of initial data. \label{item:omega2}

\item The previous two observations also hold for the ``charge
  density'' $\bar\omega:=\omega/m$ for $m=0$, where there can be no
  angular momentum. \label{item:omega2m0}

\item The angular momentum of the EMOTS shows no critical scaling. 

\item  The bound $J<M\ell$ that applies to a BTZ black hole
  formed in collapse appears to also hold for the EMOTS. It appears to
  become sharp as $p\searrow p_{*M}$ for some but not all families of initial
  data.

\setcounter{remembercounter}{\value{enumi}}
\end{enumerate}

It is hard to reconcile all this conflicting evidence. We are tempted to
dismiss the EMOTS mass as an ``epiphenomenon'' that even for $m=0$ is
somewhat gauge-dependent and not deeply coupled to the nonlinear
dynamics \cite{adscollapsepaper}. In particular, for $m>0$, the AH is
no longer constrained to be spacelike, and we have seen that it can
change shape with $p$ in a rather complicated way.  However, the key
observation for $m>0$ is simply that $p_{*M}$ is so different from
$p_*$: this seems to rule out a scenario where the same critical
solution controls both Ricci and EMOTS mass scaling.

There is no such argument for also dismissing the Ricci and $\omega$
scaling. We clearly see some threshold behaviour as $p\nearrow p_*$,
and it may be that it ends at some level of fine-tuning (or is absent
in a few families) only because the blowup associated with a critical
solution is covered up by other, non-critical, dynamics.

If we take observations \ref{item:Ricciscaling}, \ref{item:omega1},
\ref{item:omega2} and \ref{item:omega2m0} seriously, and somehow
explain observation \ref{item:noCSS} away as scaling behaviour being
``covered up'', the least implausible theoretical model appears to be
one where the dynamics as $p\nearrow p_*$ with $m>0$ is controlled by
a family of asymptotically CSS solutions, maybe having more than one
unstable mode, and admitting different angular momentum (or
``charge'') to mass ratios. A toy model for this may be the competition
between a real DSS and a complex CSS solution in a harmonic map
coupled to gravity \cite{Liebling}. 

The obvious next step is to look for these critical solutions. From
the experience with $m=0$ \cite{adscollapsepaper}, a thorough study of
asymptotically CSS solutions for $m>0$ is bound to be complex, and we
leave this to future work.


\acknowledgments 

This work was supported by the National Science Center (Narodowe
Centrum Nauki; NCN) grant DEC-2012/06/A/ST2/00397, and in part by
PL-Grid Infrastructure. This work is part of the Delta Institute for
Theoretical Physics (Delta ITP) consortium, a program of the
Netherlands Organisation for Scientific Research (Nederlandse
Organisatie voor Wetenschappelijk Onderzoek; NWO) that is funded by
the Dutch Ministry of Education, Culture and Science (Ministerie van
Onderwijs, Cultuur en Wetenschappen; OCW). JJ would like to
acknowledge financial support from Fundamenteel Onderzoek der Materie
(FOM), which is part of the NWO.


\appendix

\section{Gauge freedom}
\label{appendix:gauge}

To find the residual gauge freedom in the ansatz (\ref{trmetric}), we
define the auxiliary coordinates
\begin{equation}
\label{uvdef}
u:=t-r,\quad v:=t+r.
\end{equation}
The metric becomes
\begin{equation}
\label{uvmetric}
ds^2=-f\,du\,dv+\rb^2\left(d\theta+\beta{dv+du\over 2}\right)^2.
\end{equation}
Consider now the coordinate transformation
\begin{equation}
\label{uvgauge}
u=u(\hat u), \quad v=v(\hat v), \quad
\theta=\hat\theta+\vartheta(\hat u,\hat v).
\end{equation}
For the metric in $(\hat u,\hat v,\hat\theta)$ to retain the
form (\ref{uvmetric}), we must have
\begin{equation}
\label{betatilde}
\hat \beta=\beta{du\over d\hat u}+2\vartheta_{,\hat u}
=\beta{dv\over d\hat v}+2\vartheta_{,\hat v}.
\end{equation}
The sum and difference of these two PDEs give
\begin{equation}
\label{betatransformation}
\hat \beta={\beta\over 2}\left({du\over d\hat u}+{dv\over d\hat
  v}\right)+\vartheta_{,\hat t},
\end{equation}
where $\vartheta$ is given by 
\begin{equation}
\label{vartheta}
\vartheta(\hat t,\hat r)=\vartheta_\infty(\hat t)
+\int_\infty^{\hat r}
{\beta\over 2}\left({du\over d\hat u}-{dv\over d\hat
  v}\right)\,d\hat r'.
\end{equation}
Under the same gauge transformation, 
\begin{equation}
\label{ftransformation}
\hat f=f{du\over d\hat u}{dv\over d\hat v}.
\end{equation}
The three arbitrary functions of one variable $u(\hat u)$, $v(\hat v)$
and $\vartheta_\infty(\hat t)$ precisely parameterise the
residual gauge freedom of (\ref{uvmetric}) and hence (\ref{trmetric}).

As we shall see below in Eq.~(\ref{Awave}), modulo Eq.~(\ref{fdef}),
the metric coefficient $f$ obeys a wave equation with principal part
$f_{,uv}$. Appropriate local data for this wave equation would be the
value of $f$ on two null surfaces $(u=u_0,v>v_0)$ and
$(v=v_0,u>u_0)$. From (\ref{ftransformation}), these null data
precisely fix the functions $\hat u(u)$ and $\hat v(v)$, so $f$
(or equivalently
$A$) is pure gauge. The function $\vartheta_\infty(\hat t)$, which
parameterises a rigid time-dependent rotation of the coordinate
system, can be fixed independently by setting $\beta(t,\pi\ell/2)$.
We set
\begin{equation}
\label{gaugeBC}
\beta(\pi\ell/2,t)=0,
\end{equation}
which is the natural choice for asymptotically adS spacetimes.

\section{The Kerr-adS solution}
\label{appendix:KerradS}

Here we show that a horizon-crossing patch of the Kerr-adS metric can
be written in the form (\ref{trmetric}). 

In Schwarzschild-like coordinates $(\tb,\rb,\theta)$, the eternal exterior
Kerr-adS vacuum metric is given by
\begin{equation}
\label{KerradS}
ds^2=-\bar f\,d\tb^2+\bar f^{-1}\,d\rb^2+\rb^2(d\bar\theta+\bar\beta\,d\tb)^2,
\end{equation}
where the area radius $\rb$ is used as a coordinate, and the metric
coefficients $\bar f$ and $\bar\beta$ are given by \cite{BTZ,Carlip}
\begin{equation}
\bar f(\rb):=-M+{\rb^2\over\ell^2}+{J^2\over 4\rb^2}, \quad
\bar\beta(\rb):=-{J\over 2\rb^2}.
\end{equation}
The dimensionless mass parameter $M$ takes value $-1$ (with $J=0$) for
adS spacetime, $-1<M\le 0$ for a point particle/naked singularity and
$M>0$ for a black hole. $J$ is the angular momentum of the spacetime.
For $0<J^2/\ell^2<M^2$, $f=0$ has two roots $0<\rb_-< \rb_+<\infty$,
corresponding to an inner and outer horizon.

Defining the tortoise coordinate $\rt(\rb)$ for $\rb>\rb_+$ by
\begin{equation}
\label{dtildeR}
\rt:=\int \bar f^{-1}\,d\rb,
\end{equation}
(\ref{KerradS}) becomes
\begin{equation}
\label{TRtildemetric}
ds^2=\bar f(-d\tb^2+d\rt^2)+\rb^2(d\bar\theta+\bar\beta\,d\tb)^2,
\end{equation}
which is of the form (\ref{trmetric}), with $\rb$ now a function of
$\tb$ and $\rt$. 

In terms of the auxiliary coordinates
\begin{equation}
U:=\tb-\rt, \quad V:=\tb+\rt,
\end{equation}
both branches of the
bifurcate outer horizon can be brought to finite coordinate values
$u=0$ or $v=0$ by introducing the Kruskal coordinates
\begin{equation}
u:=-e^{-a_+U}, \quad  v:=e^{a_+V},
\end{equation}
where the constant $a_+$ is determined by the requirement that $f$ is
finite on the horizon. Further details can be found in
\cite{BTZ,Carlip}. With $t$ and $r$ then defined by (\ref{uvdef}), and
$\vartheta$ given by (\ref{vartheta}), the new metric again has the
functional form (\ref{trmetric}), but is now finite on the horizon,
with $\beta$ and $f$ given by
(\ref{betatransformation},\ref{ftransformation}).

The precise expressions for $f$, $\rb$ and $\beta$ as functions of
$(t,r)$ do not matter to us here, because in collapse simulations the
BTZ metric will not appear in this specific form, but in a generic
form related by a further {\em regular} coordinate transformation of
the form (\ref{uvgauge}). Our task is then to read off $M$ and $J$
when the BTZ metric, or a piece of it, is given in {\em generic} coordinates
of the form (\ref{trmetric}).

\section{SBP finite differencing in $r$ for the wave equation}
\label{appendix:SBP}

We assume here that the grid is centred and equally spaced, so that
$r_i=i\Delta r$ for $i=0\dots N$. We finite-difference
(\ref{Xprincipal},\ref{Vprincipal}) as
\begin{eqnarray}
&&\left(X_{,r}+\frac{p}{r}X\right)_i = \br
&&{8(\tilde X_{i+1}I_1-\tilde X_{i-1}I_{-1})-(\tilde X_{i+2}I_2-\tilde X_{i-2}I_{-2})
\over 12 \bar w_i \Delta r}, \\
&&\left(V_{,r}\right)_i =
{8(V_{i+1}-V_{i-1})-(V_{i+2}-V_{i-2})
\over 12 \Delta r},
\end{eqnarray}
where we have introduced the shorthands
\begin{equation}
I_k:=I(k,i,p):=\left(1+{k\over i}\right)^p
\end{equation}
and 
\begin{eqnarray}
\tilde X_{0}&\equiv& 0, \\
\tilde X_{1}&\equiv& \tilde v_{1}X_{1}+u_{3/2}X_{2}, \\
\tilde X_{2}&\equiv& \tilde v_{2}X_{2}+u_{3/2}X_{1}+u_{5/2}X_3, \\
\tilde X_{3}&\equiv& \tilde v_{3}X_{3}+u_{5/2}X_2, \\
\tilde X_i&\equiv& \tilde v_iX_i, \qquad i\ge 4.
\end{eqnarray}
[These formulas are equivalent to Eqs.~(57-65) of \cite{lwaveSBP}, and
  are obtained from them by cancelling a factor of $i^p$ between
  numerator and denominator.] The coefficients $u_{3/2}$, $u_{5/2}$,
$\bar v_i$ and $\bar w_i$ depend on the integer $p$ and are defined by
a 4-th order recursion relation with boundary conditions at $i=0$ and
$i\to\infty$.

While the stability of this scheme is not obvious, it is easy to see
that with $\bar v_i=\bar w_i=1$, $u_i=0$, it reduces to applying the
standard fourth-order accurate symmetric stencil to
$X'+pX/r=r^{-p}(r^pX)'$, and is therefore a fourth-order accurate
discretisation. The expression for $V'$ is, just the standard
symmetric 5-point 4-th order accurate finite difference, but it is
important to note that the method consists of both finited-difference
stencils, plus appropriate boundary conditions. 

The symmmetry boundary $r=0$ is dealt with by extending the grid to
two ghostpoints according to $X(-r)=-X(r)$ and $V(-r)=V(r)$. The outer
boundary $r_N=\ell\pi/2$ is dealt with by one-sided finite differences
for the last four grid points, namely
\begin{eqnarray}
&&\left(X_{,r}+\frac{p}{r}X\right)_{N-3,\dots,N} = \br
&& {8\tilde X_{N-5}I_{-2}-64\tilde X_{N-4}I_{-1}+59\tilde X_{N-2}I_1-3\tilde X_{N}I_3
\over 98 w_{N-3} \Delta r}, \br \\
&& {8\tilde X_{N-4}I_{-2}-59\tilde X_{N-3}I_{-1}+59\tilde X_{N-1}I_1-8\tilde X_{N}I_2
\over 86 w_{N-2} \Delta r}, \br\ \\
&& {-\tilde X_{N-2}I_{-1}+\tilde X_{N}I_1
\over 2 w_{N-1} \Delta r}, \\
&& {3\tilde X_{N-5}I_{-3}+8\tilde X_{N-2}I_{-4}-59\tilde X_{N-1}I_{-2}+48\tilde X_{N}
\over 34 w_N \Delta r}. \br
\end{eqnarray}
The corresponding one-sided finite differences for $V_{,r}$ use the
same rational coefficients, but without the weights $v_i,w_i,I_k$.

We impose the boundary conditions $\phi,\phi',V,V',X=0$ and similarly
$\psi,\psi',W,W',Y=0$ using the Olsson projection method
\cite{Olsson}, which is summarised in Appendix~G of
\cite{lwaveSBP}. This method
makes sure that the discrete energy estimate still holds after
imposing the boundary conditions. It does not matter how we discretise
the $r$-derivatives in these BCs, but we choose the $(3,8,-59,48)$
stencil for $d/dr$ at the boundary set out above.

The coefficients $\tilde v_i$ and $\tilde w_i$ are tabulated for $p=1,\dots,
22$ (thus covering $m=0,\dots, 10$), and for $i=0,\dots, 2000$ in
\cite{data}. The asymptotic
expansions
\begin{eqnarray}
\label{asympvbar}
&& \tilde{v}_i^{(1)} = 1 \br
&+& \frac{(2p-1)(p-1)p(p+1)(p+3)}{60i^4} \br
&+& \frac{(2p-3)(p-3)(p-2)(p-1)p(p+1)(p+3)}{504i^6} \br
&+& O(i^{-8}) 
\label{asympwbar}
\end{eqnarray}
and 
\begin{eqnarray}
&&\tilde{w}_i = 1 \br
&+& \frac{(2p+1)(p+1)p(p-1)(p-3)}{60i^4} \br
&+& \frac{(2p-1)(p-5)(p-3)(p-2)(p-1)p(p+1)}{504i^6} \br
&+& O(i^{-8}) 
\end{eqnarray}
are accurate to double precision arithmetic for $i>2000$ and $p\le
22$, thus complementing the tabulated values for arbitrarily large $i$.



\end{document}